\documentclass[sigconf,screen]{acmart}

\setcopyright{none} 
\acmConference[ACM CCS 2022]{ACM Conference on Computer and Communications Security}{Due 25 July 2022}{Los Angeles, TBD}
\acmYear{2022}

\usepackage{multirow}
\usepackage{booktabs} 
\usepackage{epstopdf}
\usepackage{adjustbox}

\usepackage{enumitem}

\usepackage{amsmath,amssymb,amsfonts}
\usepackage{algorithmic}
\usepackage{graphicx}
\usepackage{textcomp}
\usepackage{xcolor}
\usepackage{float}
\usepackage{pifont}
\usepackage{bbding}
\def\BibTeX{{\rm B\kern-.05em{\sc i\kern-.025em b}\kern-.08em
    T\kern-.1667em\lower.7ex\hbox{E}\kern-.125emX}}

\usepackage{caption}
\usepackage{subcaption}
\usepackage{listings}
\usepackage{color}
\usepackage{cuted}
\usepackage{xspace}
\usepackage{comment}
\usepackage{arydshln}
\usepackage{url}

\usepackage{dirtytalk}

\usepackage{hyperref}
\AtBeginDocument{\hypersetup{pdfborder={0 0 1}}}

\newcounter{numquote}
\newenvironment{lquote}{%
  \refstepcounter{numquote}%
  \quote}{\unskip~\quad\thenumquote\endquote}

\newcommand{\xxx}[1]{{\color{red} #1}}
\newcommand{\eg}{{\it e.g.,}}
\newcommand{\ie}{{\it i.e.,}}

\newcommand{\gl}[1]{\textcolor{red}{#1}}

\newcommand{\sys}{{\textit{SEmu}}\xspace}
\newcommand{\uemu}{{\textit{$\mu$Emu}}\xspace}
\newcommand{\ppim}{{P$^{2}$IM}\xspace}
\newcommand{\flag}{flags\xspace}
\newcommand{\hw}{hardware signals\xspace}
\newcommand{\ca}{C-A\xspace}
\newcommand{\sam}{SAM3\xspace}
\newcommand{\ksf}{K64F\xspace}
\newcommand{\fozt}{F103\xspace}

\definecolor{mGreen}{rgb}{0,0.6,0}
\definecolor{mGray}{rgb}{0.5,0.5,0.5}
\definecolor{mPurple}{rgb}{0.58,0,0.82}
\definecolor{backgroundColour}{rgb}{0.95,0.95,0.92}

\lstdefinestyle{CStyle}{
    commentstyle=\color{mGreen},
    keywordstyle=\color{magenta},
    stringstyle=\color{mPurple},
    basicstyle=\scriptsize\ttfamily,
    breaklines=true,
    captionpos=b,
    keepspaces=true,
    otherkeywords={status_t, ArrayObject,uint32_t},
    numbers=left,
    numbersep=1pt,
    showspaces=false,
    showstringspaces=false,
    showtabs=false,
    tabsize=2,
    language=C
}
\colorlet{punct}{red!60!black}
\definecolor{background}{HTML}{EEEEEE}
\definecolor{delim}{RGB}{20,105,176}
\colorlet{numb}{magenta!60!black}
\lstdefinelanguage{json}{
    basicstyle=\scriptsize\ttfamily,
    numbers=left,
    numberstyle=\scriptsize,
    stepnumber=1,
    numbersep=8pt,
    showstringspaces=false,
    breaklines=true,
    frame=lines,
    backgroundcolor=\color{background},
    literate=
      {:}{{{\color{punct}{:}}}}{1}
      {,}{{{\color{punct}{,}}}}{1}
      {\{}{{{\color{delim}{\{}}}}{1}
      {\}}{{{\color{delim}{\}}}}}{1}
      {[}{{{\color{delim}{[}}}}{1}
      {]}{{{\color{delim}{]}}}}{1},
}

\definecolor{codegreen}{rgb}{0,0.6,0}
\definecolor{codegray}{rgb}{0.5,0.5,0.5}
\definecolor{codepurple}{rgb}{0.58,0,0.82}
\definecolor{backcolour}{rgb}{0.95,0.95,0.88}

\AtBeginDocument{
  \setlength\abovedisplayskip{0pt}
  \setlength\belowdisplayskip{0pt}
  \setlength{\belowcaptionskip}{0pt}
  }
  \setenumerate[1]{itemsep=0pt,partopsep=0pt,parsep=\parskip,topsep=0pt}
  \setitemize[1]{itemsep=0pt,partopsep=0pt,parsep=\parskip,topsep=0pt}
  \setdescription{itemsep=0pt,partopsep=0pt,parsep=\parskip,topsep=0pt}
  \renewcommand{\paragraph}[1]{\vspace{0pt}\noindent\textbf{#1}}

\begin{document}

\title{What Your Firmware Tells You Is Not How You Should Emulate It: A Specification-Guided Approach for Firmware Emulation \\(Extended Version)}


\author{Wei Zhou$^{\dagger}$, Lan Zhang$^{\dagger}$, Le Guan$^{}$, Peng Liu$^{}$, Yuqing Zhang$^{}$}
\affiliation{
    \institution{
    }
    \country{}
}


\thanks{$^\dagger$Both authors contributed equally to this work.}






\begin{abstract}

Emulating firmware of microcontrollers is challenging due to the
lack of peripheral models.
Existing work finds out how to respond to peripheral read
operations by analyzing the target firmware.
This is problematic because the firmware sometimes does not contain
enough clues to support the emulation or even contains
misleading information (\eg~a buggy firmware).
In this work, we propose a new approach that
builds peripheral models from the peripheral specification.
Using NLP, we translate peripheral behaviors in human 
language (documented in chip manuals) into a set of 
structured condition-action rules.
By checking, executing, and chaining them
at runtime,
we can dynamically synthesize a peripheral model 
for each firmware execution. 
The extracted condition-action rules might not be complete or even be wrong.
We, therefore, propose incorporating symbolic execution to quickly
pinpoint the root cause.
This assists us in the manual correction of
the problematic rules. 
We have implemented our idea for five popular 
MCU boards spanning three different chip vendors.
Using a new edit-distance-based
algorithm to calculate trace differences,
our evaluation against a large firmware corpus confirmed
that our prototype achieves
much higher fidelity compared with state-of-the-art solutions.
Benefiting from the accurate emulation,
our emulator effectively
avoids false positives observed in existing fuzzing work.
We also designed a new dynamic analysis method to perform driver code compliance checks against the specification.
We found some non-compliance which
we later confirmed to be bugs caused by race conditions.


\end{abstract}



\maketitle

\vspace*{-2mm}
\section{Introduction}

Microcontroller units (MCU) are small, resource-constraint SoCs (system on a chip) that drive a number
of security- and safety-critical application fields such as smart homes, mobile robots, and healthcare. With
the emergence of Internet of Things (IoT) technology, security analysis (\eg~bug finding, taint analysis,
policy violation detection) of MCU-based IoT devices is becoming increasingly important. Due to the tight
coupling between firmware code and physical peripherals, IoT devices are usually tested as a whole in the real
world. However, involving real peripherals can make it significantly more difficult to achieve scalably (\eg~scale to thousands of vendors and millions of peripherals) security analysis of firmware and devices.
Because of this, in recent years,
several emulation-based techniques have been proposed~\cite{fengp2020p2im,chen2016towards,spensky2021conware,cao2020device, zhou2021automatic,johnson2021jetset,gustafson2019toward,fuzzware}.
By rehosting the firmware on a PC, many 
existing dynamic security analysis methods become possible without involving
real hardware.

\paragraph{Problems.}
A primary challenge in firmware emulation is modeling the behaviors of unknown peripherals. In particular,
when the firmware reads a peripheral register, how can we generate
an appropriate response?
A common philosophy adopted by existing work is that 
\emph{the emulator should generate responses that 
meet the expectations of firmware so that it does not crash or hang.}
In this sense, they try to learn knowledge (about unknown peripherals) from the firmware
itself. We call this line of work \emph{firmware-guided} solutions.
For example, \ppim~\cite{fengp2020p2im} automatically builds peripheral models
by observing the peripheral access patterns
when exercising the target firmware in the emulator.
Laelaps~\cite{cao2020device}, \uemu~\cite{zhou2021automatic},
Jetset~\cite{johnson2021jetset},
and Fuzzware~\cite{fuzzware} use symbolic execution to explore the 
firmware to find satisfactory response values.
However, the firmware sometimes contains {\em incomplete} or even {\em misleading} information.
For the formal, take the timing of interrupts as an example.
Existing emulators have no clue regarding which interrupt
should be delivered at a particular time.
More generally, existing methods only learn coarse-grained
static peripheral models from the firmware. 
We will shortly illustrate the missing information (not contained in firmware) and 
the negative effects with concrete examples (see Section~\ref{sec:problems}).
For the latter, if the firmware itself contains a memory bug,
existing emulators such as ~\uemu~\cite{zhou2021automatic} would try to avoid
executing code containing it, which is a wrong emulation.
As another example, \ppim reports many mis-categorizations of
registers due to the misleading access patterns, which
ultimately leads to incorrect peripheral models~\cite{fengp2020p2im}.

The incomplete and misleading information leads to serious low fidelity issues,
influencing the effectiveness, efficiency,
and applicability of firmware analysis tools that are built on top of the emulator.
\textbf{1) Effectiveness:} Driver code for complex peripherals cannot
be properly emulated.
This would exclude a large amount of real-world firmware in dynamic analyses.
For example,
none of existing work can fuzz firmware
with the Ethernet functionality.
\textbf{2) Efficiency:} Even if the target firmware can be emulated,
the execution trace may deviate significantly from that
on real hardware.
This introduces non-negligible false positives (\ie~find non-existing
bugs on infeasible paths)
and false negatives (\ie~fail to find real bugs due to missing code coverage).
Moreover, failing to deliver interrupts accurately
makes emulation progress slow (see Section~\ref{sec:problems}).
\textbf{3) Applicability:} Low-fidelity emulation also limits
the types of dynamic analysis that can be adopted.
In fact, we observe that all the related work on firmware emulation
is limited to fuzz testing~\cite{fengp2020p2im,cao2020device,zhou2021automatic,johnson2021jetset,fuzzware}
since fuzzing can naturally tolerate inaccurate emulation --
humans have to manually confirm the bugs anyway.
A high-fidelity emulator can push forward with
new security analysis applications other than fuzzing.



\paragraph{A New Direction in Automated Firmware Emulation.}
Different from existing firmware emulation solutions which are guided
by the knowledge inferred from the target firmware,
we propose \emph{specification-guided} emulation.
The specification guides the emulator in a principled way,
making emulation more accurate.
Consequently, this approach can emulate more
complex firmware, run dynamic analysis more efficiently,
and be applied to more accurate analysis tasks (\eg~specification-based
compliance check in our prototype).

\paragraph{Main Observations.}
To specify the behaviors of a peripheral circuit,
state diagrams and state tables are the \textit{de facto}
standard~\cite{statediagram}.
In fact, we frequently observe the use of finite state machines to
implement both peripheral drivers in firmware and peripheral backends in
the emulator.
Unfortunately, other than during the internal design phase,
there is no widespread adoption of any state-diagram-based peripheral
specification which can be directly interpreted by formal methods~\cite{fasano2021sok}.
On the contrary, MCU chip vendors typically release a text-based
reference manual for each chip, in which
natural language (in particular English) is used to describe peripheral behaviors.
Compared with formatted language or tables used to describe a state diagram,
the language used in chip manuals are more ad-hoc.
For example, we found many sentences similar to
``if the value of register reg\_A is equal to X, 
then the register reg\_B will be assigned with a value Y''.
On the one hand,
these ad-hoc sentences \emph{equivalently} describe the
same peripheral behaviors as a state diagram does.
In fact, it is a common practice for developers to read and 
comprehend the manuals, and then
construct state machines to implement drivers or emulator backends.
On the other hand,
it is extremely hard, if not impossible to \emph{automatically} composite
these ad-hoc sentences into a state diagram.

This work does not aim at the ambitious goal of automatically
constructing a state machine to implement peripheral backends
in emulators.
However, we have the following observations that help
us more accurately model peripherals using specifications.
1) Individual sentences
in chip manuals can be easily understood by modern
natural language processing (NLP) engines and be
formalized as some simple {\bf condition-action} (\ca) rules;
2) Without explicitly maintaining a state machine,
strategically executing these \ca rules can also emulate peripherals.

\paragraph{Proposed Solution.}
Based on the aforementioned observations,
we propose leveraging peripheral specification, in particular
chip reference manuals to model the peripheral behaviors.
The resulting model targets particular hardware which 
is independent of the emulated firmware.
This will fundamentally address the issues associated with existing
firmware-guided emulation solutions as mentioned
before (\ie~incomplete and misleading information contained in the firmware).

After obtaining enough \ca rules, at run-time,
we dynamically synthesize a
model for each peripheral accessed during the firmware execution.
Specifically, each \textbf{peripheral model} provides 
  appropriate responses on behalf of the corresponding hardware.  
In the meanwhile, we maintain the status of a peripheral by selectively
  executing certain relevant \ca rules in the right order. 
Based on our empirical study, executing these rules
implicitly maintains the corresponding state machine.
In Section~\ref{sec:intuitiveexample}, we show an intuitive example
for a UART peripheral.
The maintained status will in turn help the model select the right (subset of)  
  \ca rules when processing the next firmware request.



Due to poor documentation of the manuals
and limitations of the NLP engine itself,
we occasionally observe missing or wrong
\ca rules, leading to inaccurate peripheral modeling.
To address this problem, 
we incorporate invalidity-guided emulation proposed in \uemu~\cite{zhou2021automatic} 
to quickly diagnose the root cause.
Intuitively, if a \ca rule is missing or wrong,
it is likely that emulation would enter an invalid state.
If this happens, we use symbolic execution to quickly identify
which peripheral read operation is responsible for the
error and correspondingly fix the \ca rule.
It is worth noting that our NLP-based
approach and invalidity-guided emulation are complementary to each other.
If there exists a missing/faulty \ca rule,
invalidity-guided emulation
can help quickly find it.
But invalidity-guided emulation such as \uemu alone cannot 
address the ``incomplete and misleading information contained in firmware'' issue 
for the reasons we mentioned before.
Concretely, \uemu does not know when to issue an interrupt
and non-crashing execution cannot guarantee data correctness.





We have implemented the proposed idea
with a prototype named \sys (Specification-guided EMUlator),
and evaluated it in a systematic way.
\sys automatically extracted \ca rules from
five chip manuals for STM32F1~\cite{RMSTM32F103}, STM32F4~\cite{RMSTM32F429}, STM32L1~\cite{RMSTM32L132}, NXP K64 series~\cite{RMK64}, and Atmel SMART series~\cite{RMSAM3}, respectively,
and then we 
diagnosed and enhanced
the results with the help of invalidity-guided emulation~\cite{zhou2021automatic}.
Collectively, these peripheral models
allow us to test and evaluate the same set of firmware samples used in
existing work~\cite{fengp2020p2im,zhou2021automatic,gustafson2019toward}.
To objectively compare our solution with
existing work,
we developed a new method based on edit distance
to quantify emulation fidelity,
which measures
how an emulated execution deviates from real execution
on hardware.
Using it,
we conclude that our approach achieves much better trace fidelity compared with 
existing firmware-guided solutions.
This allows us to
emulate more complex peripheral and get better fuzzing results.
Finally, we applied our approach to a new dynamic analysis task
which we call \textit{specification-based compliance check}.
Leveraging the semantic information extracted from the manuals,
it checks whether peripheral driver implementations comply
with the logic specified in the chip manuals.
We found some non-compliance which
was later confirmed to be programming bugs.
For further research, we will open source our tools and the dataset.



To summarize, we have made the following contributes:
\begin{itemize}
	\item 
	We proposed specification-guided firmware emulation
	  to more accurately emulate firmware.
	 The core technique is to leverage  NLP techniques to automatically extract useful \ca rules from  chip manuals.
	

	\item We proposed incorporating invalidity-guided emulation
	  to  identify missing or faulty \ca rules
	  extracted by \sys.
	
	\item To quantitatively evaluate emulation fidelity,
	we proposed a new method based on modified
	edit distance that measures trace similarity.

	\item We implemented our idea and compared
	the peripheral models generated by
	\sys against
	  six ground-truth models provided by QEMU and
	  automatic models generated by existing
	  firmware-guided solutions.
	  

	\item The extracted peripheral behavior rules enabled us to 
	conduct driver code
	compliance check, and we have uncovered  
	real non-compliance bugs with it.
	

\end{itemize}

\vspace*{-2mm}
\section{background}

\subsection{Overview of MCU Peripherals}

\label{sec:peripheral}
The main functions of MCU devices are accomplished by controlling the peripherals to interact with external environments.
Therefore, the firmware running on the MCU can be considered as a collection of peripheral usages.
To properly operate a peripheral, the firmware should make sure
that the peripheral is in the right status before any access;
otherwise, unexpected behavior would occur. 
For example,
if the peripheral is busying doing hardware operations (\eg~an analog-to-digital converter, or ADC, is converting the external analogue values to the digital values), it cannot respond to firmware functions.
To communicate such important status information (\ie~
whether the peripheral is ready to respond to a certain access request),
the firmware typically first inquires about the current peripheral status
by reading the corresponding status register (\texttt{SR}), which
is memory mapped into the address space of processor (a.k.a. MMIO).
Peripherals also notify the processor (thus the firmware)
of the external events via the interrupt mechanism.
Typically, an external event causes
a state change in the peripheral.
In addition, some high-throughput peripherals like I2C, USB,
and Ethernet may use Direct Memory Access (DMA) to 
transfer the data between the RAM and peripheral without involving the CPU.



\vspace*{-2mm}

\subsection{MCU Reference Manuals}
\vspace*{-1mm}
\label{sec:manual}

As can be seen, even a simple peripheral involves
lots of hardware states.
To help firmware developers understand
peripheral behaviors, chip vendors commonly publish
reference manuals for each chip written using natural language. 
MCU reference manual is supposed to provide
essential information on how to use every peripheral,
including its registers, memory map, hardware interface, and behaviors.
Since each peripheral contains multiple registers,
the manual at least holds a section of
\textbf{register memory map} to summarize the MMIO address
for each peripheral register and its access permission (\eg~read only, write only and read/write).
Following the memory map, the functions of each register
are explicated in addition to a
\textbf{field description} table.
The field description table specifies the meaning of
individual fields in the registers, including
their names, bits, access permissions, and functions. 
In explaining the function of each field, 
it first describes how the value of the field will change.
Then, it enumerates the possible values and explains the corresponding meanings. 
Taking the UART peripheral in an NXP \textbf{K64F} as an example, 
we found the following sentence to describe
the \texttt{RDRF} field of the \texttt{SR1} register:
\begin{lquote}
RDRF is set when the number of datawords in the receive buffer is equal to or more than the number indicated by RWFIFO[RXWATER].
\label{quote:sentence}
\end{lquote} 
This sentence specifies how the field \texttt{RDRF}
will be changed to 1, which we refer to as a condition.
Based on the function,
a register field can be categorized into one of the following types.
A \textbf{status field} is used to indicate the current status of a peripheral. 
A \textbf{control field} is used by the firmware
to configure or initialize the peripheral functions.
A \textbf{data field} serves the I/O interface of the peripheral.
It is usually connected with a shift register
which is further backed by a data buffer.
While some peripherals use the same data register
to connect both the transmit buffer (for sending data)
and the receive buffer (for receiving data),
some use two dedicated data registers for
sending and receiving data.

The manual also includes
an \textbf{interrupt vector assignment} table that 
summarizes the IRQ numbers assigned for each peripheral.
Note some peripherals may have multiple IRQ numbers for different
functions.
For example, a peripheral may generate an interrupt
on error, alarm, or update.
Finally, each DMA request of a peripheral is mapped into a DMA channel.
This information is summarized in a table called
\textbf{DMA channel assignment}.
For example, the DMA request from the ADC1 peripheral of
the \fozt device is routed to channel 1,
which is activated by programming the DMA control bit of
the corresponding peripheral.
When a DMA transfer ends, the corresponding interrupt 
is triggered to notify the firmware.

As mentioned before, chip vendors commonly publish
reference manuals for developers, and
our proposed approach relies on the availability of
these manuals and the important information included in them
(\ie~register memory map, field description, etc.).
To show the prevalence of chip manuals (thus the applicability
of our approach),
we conducted a survey for
15 top players in the global MCU market~\cite{topmcu}.
The results, shown in Appendix~\ref{app:mcus},
indicate that all major chip vendors publish chip manuals and 
the manuals contain all the needed information.
It is worth mentioning that some vendors require free registration 
to access the manuals (\eg~NXP~\cite{nxp}).
Also, different vendors may use different names
for semantically similar sections.
For example, the ``register memory map section'' is called 
``memory map and register boundary addresses'' in STM32 manuals~\cite{RMSTM32F429}.
However, a vendor tends to use the same section names
across its chip lines for consistency.

On the other hand, a chip manual may not include comprehensive information
for all peripherals.
For example, in the manual of a popular BLE chip~\cite{nordic} designed by 
Nordic, 
developers are redirected to read the BlueTooth Core Specification
to fully understand the description of its proprietary BLE peripherals.
Our approach cannot support modeling these peripherals.

\vspace*{-3mm}
\subsection{Natural Language Processing}


Although reference manuals (mainly in PDF format)
are unstructured data,
there are observable characteristics in the format and text.
Hence, NLP techniques can be used to 
understand the naturally expressed sentences and
extract condition-action logic from chip manuals. 
More specifically, NLP techniques can identify the related registers and fields by recognizing the noun and verb phrases,
find the causal relationships between the conditions and actions 
through the sentence structure and conjunctions, 
and infer the semantics of the sentences based on the relation
between pairs of words.

We reuse the quoted sentence above
as an example.
It has three named entities: \texttt{RDRF}, \texttt{receive buffer}, and \texttt{RWFIFO[RXWATER]}.
The \textit{Part-of-speech (POS) tagging} technique~\cite{pos}
can be applied to recognize the named entities by identifying the characteristic structure of words (\eg~noun and verb). 
The causality can be distilled through \textit{Constituency Analysis}~\cite{cons}.
A binary parse tree is generated to divide a sentence into different constituents so that the conditions and actions can be separated.
In this example, the identified action is ``RDRF is set'' and
the remaining sub-sentence is a conditional clause.
Also, \textit{Typed Dependencies Analysis}~\cite{dep} can be 
applied to analyze the grammatical structure. 
It matches the verbs and their corresponding subjects or objects,
so that the \texttt{RDRF} sentence can be converted to first-order logic. 
Taking the sub-sentence ``RDRF is set'' as an example, \texttt{set} is a predicate that depends on the noun phrase \texttt{RDRF}.  
Since this kind of condition-action representations are
repeatedly found in chip manuals, NLP techniques can
be very effective in extracting \ca rules. 

%

\vspace*{-2mm}
\section{Motivation and Key Ideas}

We first explain how existing low-fidelity emulation solutions
impact dynamic analysis approaches, in particular fuzzing.
Then, we use an intuitive example to show
how our approach can achieve higher fidelity.

\vspace*{-3mm}
\subsection{Problems with Low-fidelity Emulation}
\label{sec:problems}

To facilitate large-scale dynamic firmware analysis, full-system emulation without any hardware dependence is essential.
Recent work has made substantial progress
in peripheral modeling, a key barrier to dynamic firmware analysis~\cite{zhou2021automatic,cao2020device,chen2016towards, fengp2020p2im,johnson2021jetset,gustafson2019toward,spensky2021conware,fuzzware}.
Different techniques have been explored recently,
such as
access-pattern recognition~\cite{fengp2020p2im},
symbolic execution~\cite{cao2020device,zhou2021automatic,johnson2021jetset,fuzzware},
and real trace analysis~\cite{gustafson2019toward,spensky2021conware}.
All these solutions share
the same philosophy -- as long as the peripheral model meets
the expectations of the firmware, the emulation is considered successful.
Here, meeting the expectations means the firmware does not crash or hang.
Therefore, they only use the firmware as the source of information to build
coarse-grained static peripheral models.
These firmware-guided solutions
only approximately emulate the firmware and thus suffer from
low fidelity issues when the input space becomes large.
Here, we  use a concrete example to explain this problem.

In Listing~\ref{lst:UART}, we show a code snippet from a real-world firmware evaluated in \ppim~\cite{fengp2020p2im}.
It runs on a PLC (Programmable Logic Controller), which continually communicates with a remote SCADA machine via the Modbus protocol over UART.
The main program logic runs inside an infinite loop.
Specifically, in each loop (called scan cycle),
the \texttt{loop()} function is invoked.
This function first checks whether there is any incoming data available  (line 3).
Only when more than seven bytes are accumulated in the buffer (line 4) will
they be fetched from the buffer and got processed (line 6).
To fill the receive buffer, the UART interrupt must
be raised by hardware so that the corresponding
handler named \texttt{UART\_IRQHandler()} will be invoked.
The handler not only receives data,
but also transmits data and handles errors.
The exact  sub-function (in the handler function) to invoke is jointly decided  by the
status register (\ie \texttt{isrflags}) and
two control registers (\ie \texttt{crlflags} and \texttt{cr3flags}).

Before presenting the problem,
we first explain how the UART hardware works together with the shown driver code in receiving external data. 
Whenever a byte is available, the hardware
moves a byte to a FIFO receive buffer inside the UART hardware,
and then sets
the \texttt{RXNE} flag of the UART status register.
If the interrupt is enabled (\ie~the flag \texttt{RXNEIE} is set),
an interrupt will also be triggered.
According to current status and control registers values, 
the handler will eventually
execute the \texttt{UART\_Receive\_IT()} function
which reads the data register.
On reading the data register, the FIFO receive buffer is shifted
so that a byte is returned via the data register to the firmware.

In this example, the approximate emulation achieved
in existing firmware-guided work exhibits at least two defects in fuzzing,
which are fundamentally caused by the incomplete information
contained in the firmware.
First, the emulator does not know the type and
timing of interrupt deliveries.
Therefore, it triggers each active interrupt  in a round-robin fashion, 
either based on the number of executed basic blocks or the consumed time.
However, there are many active interrupts during firmware execution. 
It typically takes lots of time for
the UART interrupt to be selected by the emulator.
Second, even if the UART interrupt is triggered, 
the status and control registers should hold
correct values so that the right sub-function 
(in our example, \texttt{UART\_Receive\_IT()}) can be invoked.
However, existing work cannot guarantee this since
other sub-functions (\eg~\texttt{UART\_Transmit\_IT()}) do not crash the execution either
and thus can also be selected.
Obviously, the firmware cannot tell the emulator
which sub-function should be executed -- the relevant information is not contained in the firmware.
Combined, it usually takes a long
time or relies on some non-determinisms of fuzzing
to invoke the intended function \texttt{UART\_Receive\_IT()} for receiving input.

To conclude,
firmware fundamentally lacks
some essential information for the firmware-guided emulators.
Therefore, these emulators have to blindly try all possible
combinations in the input space
(\eg~interrupt timing and status register values).
While these solutions can avoid crashes and hangs due to some clever designs,
they emphasize less on emulator fidelity.

\begin{lstlisting}[style=CStyle,label={lst:UART},xleftmargin=1em,framexleftmargin=1em,frame=shadowbox,caption={Code snippet of a real-world PLC firmware (minor modifications are made to save space).}]
int __fastcall loop(Modbus *const buffer, ...) {
    ...
    int length = HardwareSerial::available();
    if (length <= 7)
        return 0;
    Modbus::getRxBuffer(buffer);
    ...
}

void UART_IRQHandler(UART_Handle *hUART) {
    uint32_t isrflags = READ_REG(hUART->SR);
    uint32_t cr1flags = READ_REG(hUART->CR1);
    uint32_t cr3flags = READ_REG(hUART->CR3);
    uint32_t errorflags = (isrflags & ...)
    /* UART in Receiver mode */
    if((isrflags & UART_SR_RXNE) != RESET && ...){
        UART_Receive_IT(hUART);
        ...
    }
    /* UART in Transmitter mode */
    if((isrflags & UART_SR_TXE)) && ...){
        UART_Transmit_IT(hUART);
        ...
    }
    errorflags = (isrflags & (UART_SR_PE | UART_SR_FE | UART_SR_ORE | UART_SR_NE));    
    /* UART in Error mode */
    if((errorflags != RESET) && ...)
        ...
    ...
}
\end{lstlisting}

\begin{figure}[t]
\setlength\abovedisplayskip{0pt}
\setlength\belowdisplayskip{0pt}
\centering
\includegraphics[width=0.85\columnwidth]{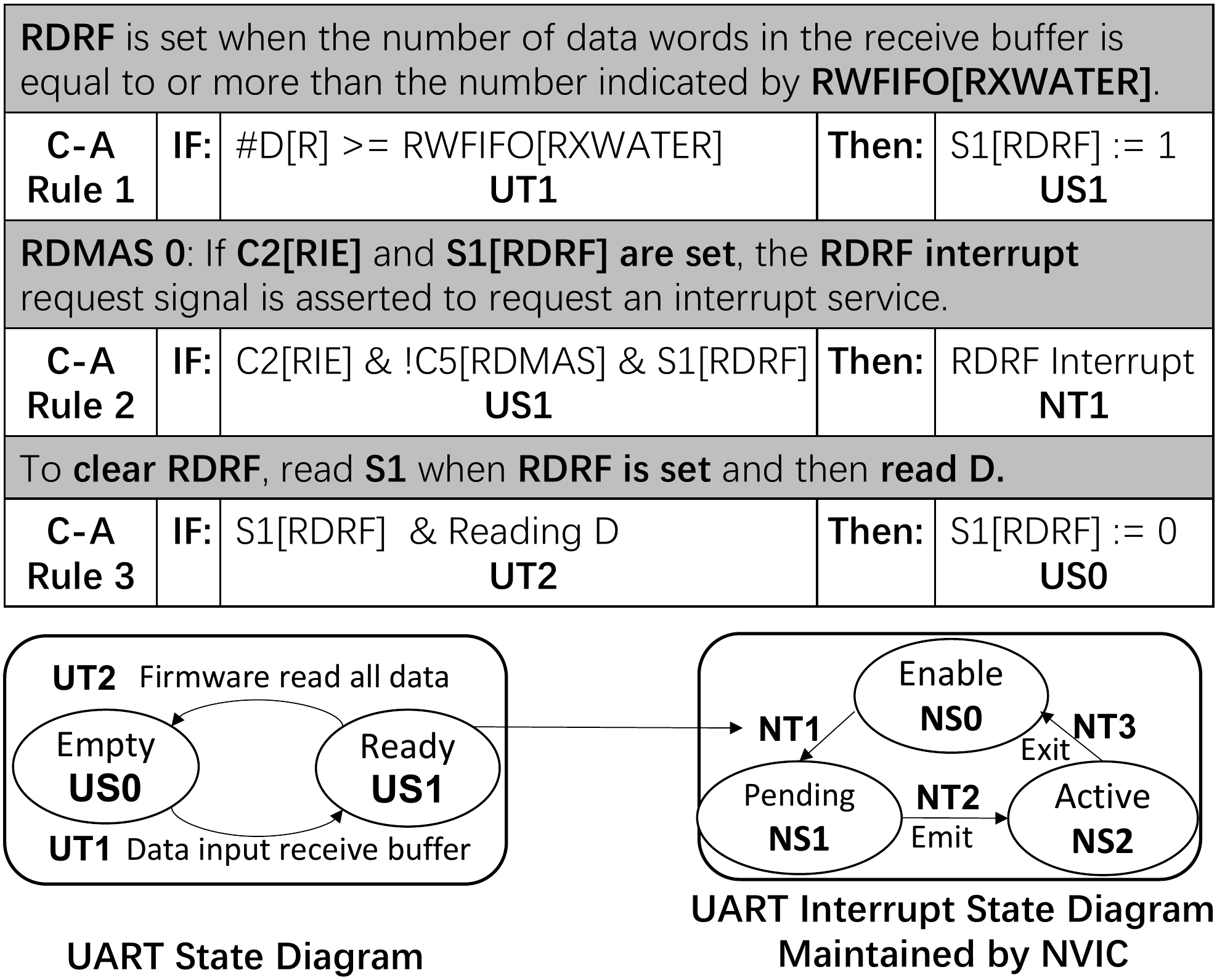}
\vspace{-4mm}
\caption{Checking, executing, and chaining condition-action rules can emulate state transitions.}
\flushleft
\scriptsize{
Note: In the sentence describing \ca rule 2,
``RDMAS 0'' indicates the following sentence is valid when the RDMAS field is 0. 
}
\label{fig:statemachine}
\vspace{-6mm}
\end{figure}

\subsection{Using \ca Rules to Emulate State Transitions}
\label{sec:intuitiveexample}

To address the low-fidelity problem, our approach
automatically extracts structured \ca rules from chip manuals
and strategically executes them.
This process emulates
the state transitions of the underlying hardware without
\emph{explicitly} maintaining the state machine,
making the emulation more principled
and thus yielding good fidelity.
We use Figure~\ref{fig:statemachine} 
to illustrate the idea using the UART peripheral as an example.
The top table lists three sentences excerpted from the original
reference manual for K64F series chips~\cite{RMK64}
and \ca rules extracted from them.
We highlight the main subjects/objects in these sentences.
In our NLP engine, they are \textbf{named entities}
against which a condition or action is specified.
At the bottom, we show the state diagrams that
we manually constructed by reading the manual.
We note this is an essential step for developers
to implement the peripheral drivers traditionally.
\ca rule 1 states that if the condition
``\#D[R] >= RWFIFO[RXWATER]'' is satisfied (UT1),
then an action ``S1[RDRF] := 1'' will be invoked (US1).
Note how this \ca rule corresponds to the edge
UT1 and node US1 of the UART state diagram.
The action assigns value 1 to the register field \texttt{S1[RDRF]},
which in turn makes the condition of \ca rule 2 true (US1).
A UART interrupt is generated as a result of the corresponding action (NT1).
It makes the UART interrupt pending (NS1).
Note this action has a cross-peripheral effect that influences
the state machine of NVIC,
Arm's built-in interrupt management peripheral.
The \ca rule 3 is related to receiving data.
How it maps the state transitions is self-explanatory.
As can be seen, by simply checking, executing, and chaining each \ca rule,
we can achieve the same results of
the state transitions governed by the state diagrams that we constructed manually.


\begin{figure*}[t!]
\setlength\abovedisplayskip{0pt}
\setlength\belowdisplayskip{0pt}
\centering
\includegraphics[width=.6\textwidth]{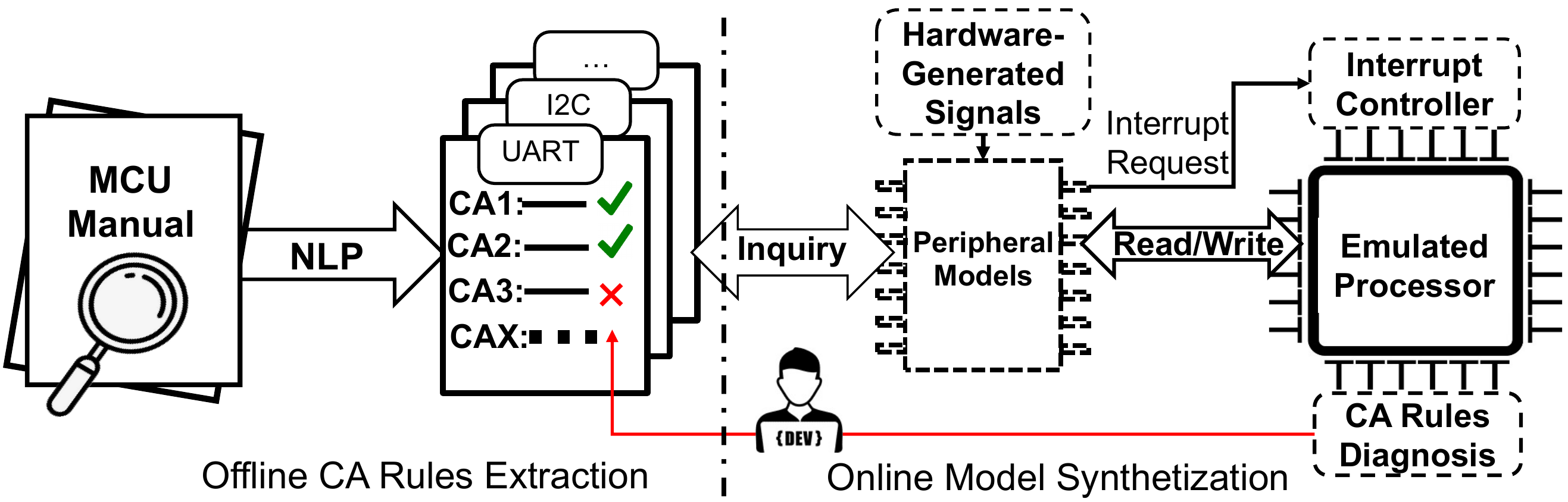}
\vspace{-3mm}
\caption{\sys overview}
\label{fig:arch}
\vspace{-5mm}
\end{figure*}


\vspace*{-2mm}
\section{Overview}

We leverage peripheral specification to guide the emulator to overcome the aforementioned 
limitations of previous work.
In a sense, we mimic a real-world emulator development process
in which the developers read the chip manuals and
correspondingly write peripheral models as the emulator backends.
However, to make it scalable,
we adopted a semi-automatic method by leveraging the latest
advances in NLP. 
As shown in Figure~\ref{fig:arch},
given a chip manual, we use an NLP engine to extract
a set of \textbf{Condition-Action (\ca)} rules that describe peripheral behaviors (Section~\ref{sec:carule}).
The key observation is that while the natural languages used in chip manuals 
are diverse,
most sentences are similarly structured, making
the extraction of \ca rule accurate enough.
With these rules, at run-time, the emulator intercepts
the firmware-peripheral
interactions, which drive the checking, executing and
chaining of \ca rules. 
As mentioned before, this emulates state transitions of peripheral
hardware without explicitly maintaining the state machine.
When firmware issues a read request to a peripheral register, 
the emulator inquires the current peripheral status and calculates a response.
Therefore, our approach dynamically builds a peripheral model
according to the specification, achieving \emph{state-aware} firmware
emulation (Section~\ref{sec:modelSynthesizing}). 

Due to imperfect documentation and limitation of the NLP engine itself,
we occasionally observed failed emulation.
Therefore, we also developed a symbolic-execution-aided
diagnosis technique to quickly pinpoint the faulty or missing \ca rules.
The diagnosis results are manually analyzed by human analysts
who then revise the automatically extracted \ca rules to fix the
problem (Section~\ref{sec:invalid}). 


On top of the proposed state-aware emulator,
we developed two security analysis plugins.
First, an AFL-based fuzzer is developed to find memory-related bugs in firmware.
Second, we develop a compliance check tool that 
tracks the MMIO access sequence for each peripheral
and compares it with the static rules in specification.
Unless otherwise specified,
we use peripheral descriptions from
the NXP K64F manual~\cite{RMK64} as examples to illustrate ideas.

\vspace*{-2mm}
\section{Extracting \ca Rules from Chip Manuals}
\label{sec:carule}

In this section, 
we first introduce the formal definition of 
conditions and actions, along with their classifications.
Then, we elaborate on how to identify conditional clauses
and extract trigger-action rules.

\vspace*{-3mm}
\subsection{Conditions and Actions}
\label{sec:cadef}

The condition-action rule is the most critical concept in our system.
It describes how important events such as accessing
an MMIO register influence the peripheral state.
A {\em condition} comprises one or more predicates combined 
using the Boolean $and$ operator.
An {\em action} is one or more assignment functions. 
A \ca rule connects one condition and one action --
when the condition is met, the paired action will be taken. 
If multiple conditions are satisfied, all the associated actions
should be executed.
The subjects/objects in a \ca rule are modeled
as named entities (\eg~a register field).
Therefore,
each predicate in a condition 
specifies whether the value of a named entity 
is equal to, greater than, or less than a reference value or 
the value of another named entity. 
Each assignment function assigns a value to a named entity.  
Formally, a \ca rule is represented as the following logical expression: 
$P_1 \wedge  P_2 \wedge ... \wedge P_n  \rightarrow F_1,F_2,...,F_m$,
where $P$ refers to a predicate and $F$ refers to an assignment function.

Conditions are event-driven.
That is, when certain events happen,
a condition might be satisfied.
We group conditions based on the relevant named entities and firmware operations.
Events are delivered via signals.
Based on how the underlying 
signal is generated, we classify the affected
conditions into three types.
{\bf Type-1 condition (via external-hardware-generated signals):} 
These signals are driven by external hardware events, 
such as receiving new data from the physical UART interface.
The result can often be abstracted
as filling an internal buffer with new data.
The description for this type of 
conditions usually contains a keyword ``hardware'' or ``buffer''.
In the quoted sentence of Section~\ref{sec:manual},
``the number of data words in receive buffer'' which
is the subject of the condition,
is related to a hardware-generated signal.
{\bf Type-2 condition (via firmware-generated signals)}:
This type of conditions is driven
by firmware operations. 
For example, in the sentence ``LBKDIF is cleared by writing a 1 to it'', the write operation 
is a signal to update the value of \texttt{LBKDIF}.
{\bf Type-3 condition (via internal signals)}:
When a register value has been updated by due to
the execution of any previous actions,
some related conditions may become true.
We call this chained \ca rule execution.
Therefore, each relevant
\ca rules are checked when a previous action causes a register update.
We frequently found interrupt delivery or DMA request
as a result of chained \ca rule execution.
The \ca rule 2 in Figure~\ref{fig:statemachine} shows such an example.




Depending on how the firmware interacts with the peripheral (\ie~MMIO access
and interrupts/DMA requests),
there are three types of actions. 
{\bf Type-1 action (MMIO register related):} 
These actions update the value of a register field.
In the previous example, the action is ``RDRF is set'', which
sets 1 to the \texttt{RDRF} bit of register \texttt{S1}.
{\bf Type-2 action (interrupt related):} 
These actions send an interrupt request to the interrupt controller.
{\bf Type-3 action (DMA related):}
These actions generate a DMA transfer request
or an interrupt request when a DMA transfer is completed.
These two types of actions are normally triggered by 
buffer-related signals. 
For example, the sentence describing
\ca rule 2 in Figure~\ref{fig:statemachine} elucidates a \ca rule
containing a type-2 action;
meanwhile, when the \texttt{RDMAS} field holds 1 as indicated in the manual, the sentence,
``RDMAS 1: if C2[RIE] and S1[RDRF] are set, the RDRF DMA request signal is asserted to request a DMA transfer'' 
will generate a type-3 action.




\vspace*{-2mm}
\subsection{\ca Rule Extraction}
\label{sec:caext}

There are several challenges to extract 
the relevant \ca rules automatically:
1) How can we identify the sentences which are associated with \ca rules?
2) How can we recognize and handle co-references which are very common?
3) How can we identify the causal relationship of the sentences?

\begin{figure}[t]
\setlength\abovedisplayskip{0pt}
\setlength\belowdisplayskip{0pt}
\centering
\includegraphics[width=0.9\columnwidth]{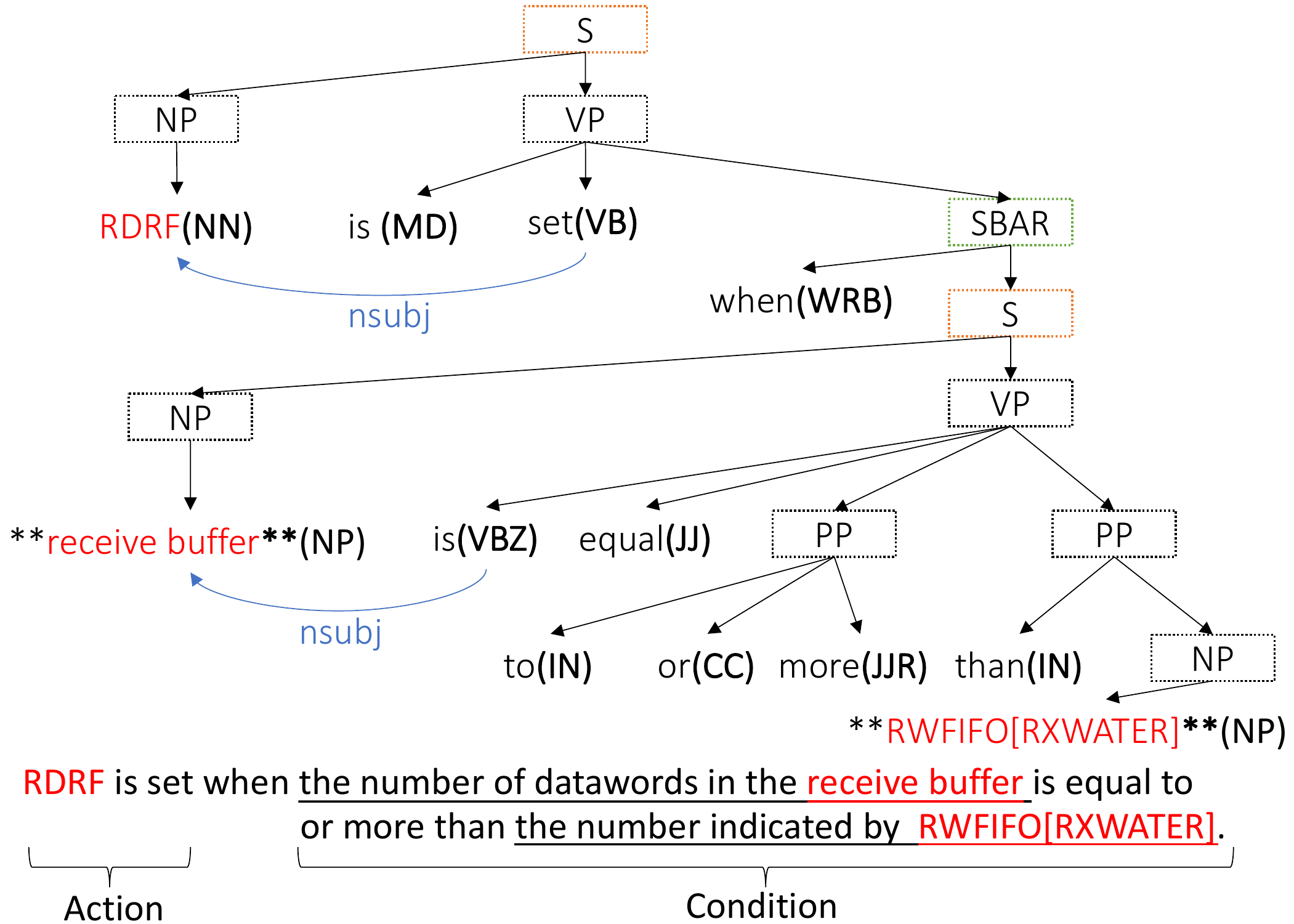}
\flushleft
\scriptsize{
1. Bold characters are the parts of speech (POS)~\cite{pos} (\eg~\texttt{NN} for nouns).\\
2. Blue edges label the dependencies between two words.
(\eg~\texttt{nsubj} indicates the dependency between a predicate and a subject). \\
3. Red characters are named entities.\\
4. **Phrase** is in abbreviated form. The full phrase is underlined in the sentence.\\}
\vspace{-3mm}
\caption{Constituency analysis of the example sentence of Section~\ref{sec:manual}.}
\label{fig:nlp_example}
\vspace{-5mm}
\end{figure}


\vspace*{-2mm}
\subsubsection{Collecting Relevant Sentences via Named Entity}

A chip manual contains thousands of sentences, many of which 
are not related to any meaningful \ca rules and
should be filtered out.
To collect relevant sentences, we take advantage of named entities.
As mentioned before, the subjects/objects in \ca rules
are represented as named entities. 
We first identify a set of important named entities.
Based on them, we directly match the occurrences in a sentence
to collect relevant sentences.
We consider two sources of named entities.
First,
firmware interacts with peripherals via the MMIO registers.
Therefore, peripheral registers and their fields 
are a major source.
Second, we also consider the receive/transmit buffers which are
connected with data registers as named entities.


The initial set of named entities is extracted from the \textbf{register memory map} section of the manual. 
This is straightforward because this section is the central
place where registers are enumerated.
Then we scan the \textbf{fields description} part
to find relevant sentences that contain at least one named entity.
During this process, we extend the set of named entities
with newly encountered subjects/objects, such as Ethernet transmit descriptors (TDES0). 
With the new entities, we also extend the search scope
to other sections in the manual such as 
\textit{functional description} and \textit{application information}.
As new named entities and relevant sentences are found,
we will re-run the search algorithm over the already scanned
sections. 
This is an iterative process which ends
when no new named entities or sentences can be found,
or exceeding a threshold (three by default in our prototype).

 


\vspace*{-2mm}
\subsubsection{Identifying Co\-references}\label{sec::coreference}

A well-known challenge in NLP is that multiple expressions
in the natural language can refer to the same entity.
For example, we have encountered at least the following
different expressions
for the UART status register:
status register of UART, \texttt{UARTx\_SR}, and \texttt{SR}.
To uniformly represent them in \ca rules,
we have to identify all these co-references.
We address this issue by matching the random noun phrases
with the initial set of named entities extracted 
from the \textbf{register memory map},
because noun phrases in a sentence are often related to named entities.
Noun phrases are marked
as ``NP'' in \textit{Stanford POS Tagger}~\cite{pos}.
For instance, 
the underlined phrase ``the number indicated by RWFIFO[RXWATER]''
in Figure~\ref{fig:nlp_example} is a noun phrase.
We use approximate string matching to measure the similarity
between an unknown noun phrase with each of the initial set of named entities
to identify possible co-references.
In this example, the word ``RWFIFO[RXWATER]'' is close to
``UART FIFO Receive Watermark(UARTx\_RWFIFO)''. 
Thus, we know ``RWFIFO'' is a synonym of the register with the named entity
``UARTx\_RWFIFO''. 
By further searching the field description of the register
\texttt{UARTx\_RWFIFO},
we identify ``RXWATER'' as a field of this register.


\vspace*{-2mm}
\subsubsection{Identifying Conditions and Actions}

Given a sentence, we then apply
\textit{Stanford Constituency Parser}~\cite{cons},
a popular language model,
to analyze the grammatical structures of sentences. 
The conditional clauses (in the sentence) can be identified through constituency analysis 
as shown in Figure~\ref{fig:nlp_example} which uses
the quoted sentence in Section~\ref{sec:manual} as an example.
Here, a sentence ``S'' is defined as a noun phrase ``NP'' followed by a verb phrase ``VP''.
The sub-tree with ``SBAR'' as the root node stands for a 
subordinate conditional clause starting with the conjunction ``When''.  After 
identifying the conditional clause, the sentence is divided into two 
sub-sentences indicating the condition and the action, respectively. 
Due to the complexity of natural language, sometimes the Stanford parser 
cannot identify complicated conditional clauses. 
For example, the Stanford parser cannot
recognize the conditional clause in
the sentence ``LBKDIF is cleared by writing a 1 to it'',
To address this issue, we extend the grammar patterns used by the parser to specify 
syntactic constituents. 
Grammar patterns are a series of regular expressions that divide a sentence into several clauses. In the above example, 
two sub-sentences, $<NP, VBZ, VBN>$ and $<VBG, NP, IN, NP>$, are matched. 
Here, $VB.*$ represents different verb forms, such as gerund/present participle ($VBG$), present tense ($VBZ$), and past participle($VBN$).
$IN$ matches all preposition/subordinating conjunctions (\eg~in, of, to).
The second sub-sentence ``writing a 1 to it'' is a prepositional
phrase for the first sub-sentence ``LBKDIF is cleared''. 
So, we assign the second sub-sentence as the condition and the first one as the action. 
\vspace*{-2mm}
\subsubsection{\ca Rule Representation} 
\label{rulerepresentation}

After fully ``understanding'' the sentences,
our NLP engine outputs a formal representation
of \ca rules for the online model synthetization module.
We first translate conditions and actions into predicates and assignment functions.
To do so,
we employ the \textit{Stanford Dependency Parser}~\cite{dep} to connect a named entity with the related verbs.
As shown in Figure~\ref{fig:nlp_example}, 
the verb ``set'' is connected to the named entity \texttt{RDRF}.
Then, the value and operator semantics involved in the verb phrase need to be extracted. 
For this purpose, we build a mapping table,  
in which one or more keywords are mapped to a particular 
binary value (\eg~keyword ``set'' is mapped to the
value \texttt{1} and ``cleared'' is mapped to \texttt{0})) 
or Boolean operators (\eg~the Boolean operator ``equal to or more than'' is mapped to \texttt{>=} and the
assignment function is mapped to \texttt{:=}). 
Note the construction of the mapping table
requires domain knowledge but is a one-time effort.

Second, we formulate
rule triggers and actions.
Rule triggers are recognized by parsing 
condition descriptions.
For instance, 
for the sentence ``the \texttt{LBKDIF} field is cleared when firmware writes 1 to it'', 
we should check the corresponding \ca rule only when
the firmware writes to this field. 
We define five types of triggers.
1) \texttt{B-triggers} represent the conditions which are checked when data in receive/transmit buffer has changed; 
2) \texttt{W-triggers} represent the conditions which are checked on firmware write operations; 
3) \texttt{R-triggers} represent the conditions which are checked on firmware read operations; 
4) \texttt{V-triggers} represent the conditions which are checked when the value of a field is updated by internal signals;
5) \texttt{O-triggers} represent the conditions which are checked on any other signals such as the timer or manual invocations.
If one rule comprises multiple conditions, we use the \texttt{\&} symbol to concatenate them.  
It is straightforward to recognize actions from the
parsed sentences. They are encoded depending
on their types mentioned in Section~\ref{sec:cadef}.

Lastly, we formulate the \ca rules.
The named entities are formally represented as 
\texttt{Reg[Field]} (\eg~\texttt{RWFIFO[RXWATER]}).
For data registers, we 
use \texttt{D[R]} and \texttt{D[T]} to represent the 
corresponding receiver and transmit buffers. 
\texttt{\#} at the beginning of each buffer-related keyword indicates the current occupied size of that buffer.
We use the $\rightarrow$ symbol to
separate the conditions and actions in a \ca rule.
Formal representation of the resulting \ca rules are presented
in Appendix~\ref{app:carule}.
We use the first sentence in Figure~\ref{fig:statemachine}
to exemplify how a sentence is formulated.
The translated \ca rule is:
\begin{equation}\label{equ:cond5}
\setlength\abovedisplayskip{1pt}
\setlength\belowdisplayskip{1pt}
    B \quad \#D[R] \geq RWFIFO[RXWATER] \rightarrow S1[RDRF]:= 1\nonumber
\end{equation}
``B'' indicates that this \ca rule is triggered by a B-trigger.
The phrase 
``the number of datawords in the receive buffer'' is formalized as \texttt{\#D[R]}, which should be ``equal to or more than''
RWFIFO[RXWATER].
The latter entity is directly extracted from the sentence
and it already meets the required \texttt{Reg[Field]} format.
The action is to assign 1 to ``RDRF'',
which our parser automatically finds its holder register ``S1'',
yielding an action ``S1[RDRF] := 1''.
We show all the extracted \ca rules for K64F UART in Appendix~\ref{app:uartrule}.

\vspace*{-2mm}
\section{Synthesizing Peripheral Models with \ca Rules}
\label{sec:modelSynthesizing} 






We use QEMU to emulate the basic ARM ISA and
core peripherals (\eg~NVIC).
During firmware emulation, 
we dynamically build a model for each peripheral,
taking the intercepted
firmware-peripheral interactions 
and the extracted \ca rules as inputs.
By capturing the rule triggers mentioned
in Section~\ref{rulerepresentation},
we check whether the condition of any \ca rule becomes satisfied.
Since the intercepted interactions only contain
the raw addresses (\eg~reading/writing a value to an address),
we have to first translate the target addresses into named entities,
which is trivial because the \textbf{register memory map}
contains the needed information.
If the named entity matches a predicate and
the condition is satisfied, the corresponding action will be executed.
We initialize a data structure of each peripheral,
which contains 1) the current values for the affiliated
named entities, and 
2) the states of interrupt and DMA request (if any).

\label{sec:model}

As mentioned in Section~\ref{sec:cadef},
there are three types of conditions.
They may become satisfied in the presence of triggers from
external hardware, firmware interaction, or internal hardware.
Among them, the firmware interactions can be directly captured
by intercepting the MMIO accesses.
Internal hardware triggers come from the results
of previous \ca executions.
Specifically, the outcome of a \ca rule execution
triggers another \ca rule execution,
which in turn can trigger another one.
In essence, this enables \textbf{chained execution} of
\ca rules, a very important aspect in emulating
peripheral behaviors.

To capture triggers from \emph{external} hardware,
we observe that most \hw are related to data transmission.
This allows us to model \hw by monitoring the I/O
interface, in particular the transmit and receive buffers.
For example, 
the \texttt{RDRF} field in UART 
is set when the receive buffer is full.
Since the hardware receive/transmit buffers are used for external
I/O channel, 
we emulate these buffers with two byte arrays.
When the firmware writes a byte to the data register,
we move it to the transmit buffer.
When the firmware read from the data register,
we return a byte from the receive buffer like real hardware.
Consequently, buffer-related conditions like the number of available bytes in 
a receive or transmit buffer can be easily emulated. 
Likewise, we can also emulate timer-based hardware signals with software. 
For the rest of few \hw which cannot be emulated,
we keep the corresponding conditions to be always satisfied.
This makes the associated actions
get a fair chance to be invoked.
For example, the \texttt{SR[STRT]} field of
the ADC peripheral in F103 
is set by hardware when regular channel conversion starts.
Since we have no clue as to when the conversion would start,
we set the \texttt{SR[STRT]} field to be always one.
While this may influence emulation fidelity,
we found it helpful in pushing emulation forward.

\vspace*{-3mm}
\section{Diagnosing Faulty/Missing \ca Rules}
\label{sec:invalid}

The extracted \ca rules may be incomplete or incorrect due to two reasons. 
First, as mentioned before,
some hardware signals cannot be emulated.
We use a workaround that sets the corresponding condition
to be always true, leading to the incorrect execution of certain \ca rules.
Second, state-of-the-art NLP techniques 
face difficulty in handling very complex sentences such as
those with underlying or nested conditions.
Faulty or missing \ca rules influence emulation fidelity
and thus must be minimized.
However, it would involve substantial human efforts in locating
the root cause.

To reduce the needed manual efforts, 
we propose an invalid-state-detection-based \ca rule checker to automatically diagnose the faulty/missing \ca rules.
Inspired by \uemu~\cite{zhou2021automatic},
we find symbolic execution very good at reasoning
about the root cause of failed emulation.
Failed emulation typically leads to
an invalid execution state (\eg~stall or crash),
and symbolic execution can help us quickly locate the
improper response for the peripheral reading.
Concretely, for the target peripheral,
we first prepare a firmware sample and a valid testcase
that correctly executes with the firmware.
This can be verified on real devices.
Then we run this sample on \sys and collect the concrete responses
generated by the synthesized peripheral models.
These concrete responses are fed to a symbolic execution engine
and used to guide symbolic execution
in selecting branches,
similar to the concolic mode implemented in S2E~\cite{concolic}.
If an invalid state is detected, it must be due to wrong \ca rules,
because we have specifically selected a firmware sample and testcase
that do not trigger any invalid state.
Since an invalid state is encountered, 
one of previous responses from our peripheral model must be wrong.
We will take the other branch in the last conditional
statement and solve the corresponding symbolic variable.
This symbolic variable tells us the address of the wrongly modeled register.
We further compare the wrong value generated by our model with
the one solved by symbolic execution engine
to identify the incorrect bits.
Finally, our tool lists all the executed \ca rules
related to this field in reverse order.
Humans must now be involved to confirm the wrong rule.
Diagnosis is an iterative process.
In each round, it fixes one \ca rule until the used firmware can
be emulated correctly.

Taking the description of
\texttt{SR[MOSCXTS]} for the PMC peripheral of
the SAM3X chip as an example,
our NLP engine failed to generate any \ca rule for this field at first. So our model used the reset value (zero) by default.
However, during PMC initialization, the diagnosis tool detected an
invalid state due to a wrong response from 
reading the \texttt{SR} register.
In particular, our diagnosis tool indicated that
the response should be \texttt{0x1} at that point.
Checking the manual, it does not clearly state that
this field depends on a hardware signal (\ie~no ``hardware'' keyword
in the description) 
and no conditional clause was found.
We addressed this issue by adding a static rule that 
\texttt{SR[MOSCXTS]} should always be set.

\vspace*{-2mm}
\section{Evaluation}

We have implemented the proposed ideas in a prototype
called \sys.
To evaluate it, we aim to answer the following research questions.  
Q1) Can our NLP engine automatically extract \ca rules to
describe peripheral behaviors?
Q2) Can the diagnosis tool help correct incorrect rules?
Q3) Are the extracted \ca rules complete and sound?
Q4) Can peripheral models dynamically built with \ca rules
provide higher fidelity compared with firmware-guided approaches?
Q5) Will the improved emulation fidelity provide better performance? 
Q6) Can compliance check which requires more accurate peripheral modeling
help us find bugs?


We selected five chip manuals that cover 
more than twenty popular MCU series including
STM32\textbf{F103}~\cite{RMSTM32F103}, STM32\textbf{F429}~\cite{RMSTM32F429}, STM32\textbf{L152}~\cite{RMSTM32L132}, NXP \textbf{K64F} series~\cite{RMK64} and Atmel \textbf{SAM3X} series~\cite{RMSAM3} belonging to three top MCU vendors.
It is worth noting that a manual can
cover a series of MCU chips that share similar peripherals.
For example, STM32\textbf{F429}~\cite{RMSTM32F429} also covers
STM32F405/415, STM32F407/417, STM32F427/437 
and STM32F439 series MCUs.
Supporting these MCU models
allows us to test and evaluate the same set of firmware samples used in
existing work~\cite{fengp2020p2im,zhou2021automatic,gustafson2019toward}.
We first extracted initial \ca rules from these manuals and
evaluated if they can support emulating a set of unit-test samples (Section~\ref{sec:extraction}).
We then conducted \ca rule diagnoses to improve the \ca rules and
used the enhanced rules to test the same unit-tests 
and a set of real-world firmware and 
complex demo programs shipped with chip SDKs which cannot be handled  in previous work
(Section~\ref{sec:enhance}).
To explain why the enhanced \ca rules include enough information
for building peripheral models,
we evaluated their faithfulness to the ``ground-truth'' used in
QEMU peripheral models (Section~\ref{sec:faithfulness}).
Using the proposed method to measure trace similarity, we further conducted 
emulation fidelity tests (Section~\ref{sec:fidelity}),
and fuzzed real-world firmware samples (Section~\ref{sec:fuzz}). 
Finally, we performed the compliance check between
the driver implementation and the
specification (Section~\ref{sec:comp}).
All the experiments were conducted on
a 16-core Intel Xeon Silver 4110 CPU @ 2.10GHz 
server with 48 GB DRAM running a Ubuntu 18.04 OS.

\vspace*{-2mm}
\subsection{\ca Rule Extraction}\label{sec:extraction}



For each manual,
we extracted \ca rules for 26 popular peripherals, including ADC, 
I2C, SPI, GPIO, UART, Ethernet, etc.
In Table~\ref{tab:nlp},
we show the statistics of raw \ca rules obtained via our NLP engine.
For each peripheral, we list the number of involved sentences  (\textbf{\#Sent.}), registers and fields (\textbf{\#Reg(Field)}).
For the extracted \ca rules, we broke
them down based on their categorization information
mentioned in Section~\ref{sec:cadef}.
The last column shows the number of total rules
and the number of revised rules if any.
In total, we extracted
3,602 unique \ca rules from 23,424 
sentences involving 26 different peripherals.

As shown in the Table~\ref{tab:nlp},
more than half of the conditions are
triggered by firmware-generated signals (C2).
This indicates that MMIO interaction contributes the most
to peripheral operations.
15.8\% of conditions are associated with hardware-generated signals (C1), including the I/O interactions and other hardware signals.
The rest (20.9\%) are triggered by internal conditions (C3).
Most of them are related to interrupt or DMA requests
that become active as a result of another \ca rule execution (\ie~chained execution).
For actions,
84.2\% of \ca rules hold type-1 action (A1) to update the state of fields.  
More than 14.2\% of \ca rules are designed to generate interrupts (A2)
and 1.57\% of rules 
are used to send DMA requests (A3).
In general, complex peripherals need more \ca rules,
as exemplified by the Ethernet and PWM peripherals.
The same type of peripheral on different MCU SoCs
may exhibit different statistics, due to the 
different implementation and presentation flavors.
For example, on SAM3X,
the state information about interrupts, 
access permission and channel modes are separately
maintained in three registers,
while K64F and STM32 combine them in one register.
Moreover, the number of \ca rules for GPIO on SAM3X
is noticeably higher than others.
We found that this is because the GPIO lines on SAM3X are 
managed by the PIO controller, 
which requires complex configuration via 
32 programmable input/output lines with three registers.

\paragraph{Using the Raw \ca Rules for Firmware Evaluation.}
We used the raw \ca rules to build peripheral models for 66 unit-test samples 
released in the \ppim paper~\cite{fengp2020p2im}. 
These unit-tests run on three chips,
covering various peripherals and OS libraries.
For these 66 test cases, \ppim and \uemu
achieve 83\% and 95\% passing rates, respectively.
Unfortunately, during firmware booting, our emulator
failed to correctly run the clock configuration for
these unit tests.
Specifically, the raw \ca rules for RCC on STM32F103, MCG on K64 and PMC on SAM3X cannot
faithfully to build usable models for these peripherals.
Since these peripherals provide clock sources
that are on the non-bypassable booting bath,
the models built from raw \ca rules failed all the unit tests.
However, regarding individual peripherals,
the raw \ca rules
work correctly for most other peripherals without enhancement,
as indicated by the last column of Table~\ref{tab:nlp}.
After revising clock related rules, our method achieved a passing rate of 96.97\%.
Only two F103 I2C unit tests cannot pass.
We explain the failure reasons and how we semi-automatically revise the rules
for clock peripherals in Section~\ref{sec:enhance}.


\vspace*{-2mm}
\subsection{\ca Rule Enhancement}
\label{sec:enhance}


As shown before, when using the raw \ca rules to synthesize peripheral models,
we encountered failed emulations. This lies in the limitation of NLP
techniques and unavoidable ambiguous or even wrong hardware 
descriptions\footnote{Some typos have been observed in the documents.
We have reported to the relevant maintainers}.
While we cannot
fundamentally solve this problem, a symbolic-execution-aided diagnostic tool
has been developed to quickly find out which rule is incorrect. Then we can
manually enhance the rules. Following the descriptions mentioned in
Section~\ref{sec:invalid}, we first prepared 61 firmware samples that
demonstrate individual peripheral usages in different working modes
(\eg~polling, interrupt and DMA mode for UART) from vendor SDKs. We then ran
the samples with \sys and used the collected concrete responses to guide
symbolic execution. If the symbolic execution engine enters an invalid state,
it terminates and outputs diagnosis information including the involved
registers and \ca rules. Note that when an invalid state is detected, it must
be due to wrong \ca rules because we have made sure the tested samples and
test-cases do not crash on real devices. In total, we added 26 rules
(0.6\%) and fixed 21 rules (0.5\%) as shown in the brackets in the last
columns of Table~\ref{tab:nlp}. To be specific, we added
5 \ca rules to I2C (F103, F429 and L152, 15 rules in total),  3 rules to ADC DMA mode (SAM3X), 3 rules for Ethernet (F429), 4 rules for MCG (K64), and 1 rule for PMC
(SMA3); we also modified 6 rules for RCC (F103, F429 and L152, 18 rules in total) and remove 3
useless rules for SPI debug mode (K64).



We use RCC, the clock peripheral for STM32F103 to explain how incorrect rules
were corrected with the help of our diagnostic tool. The original sentence to
describe the \texttt{CFGR[SWS]} field of RCC is ``set and cleared by hardware
to indicate which clock source is used as system clock.''. Since the NLP
engine does not know how hardware will set/clear the field, our tool
generated a rule \underline{$O \quad * \rightarrow CFGR[SWS] := 0/1/2/3$}, which indicates that 1) this
is an O-trigger (Section~\ref{rulerepresentation}), 2) the condition is
always satisfied, and 3) the action is to set \texttt{CFGR[SWS]} with a
random value from 0-3 (since this is a 2-bit field). When we used this rule
against the firmware sample, we found that the RCC driver always failed
booting. Using the diagnostic tool, we were able to capture the invalid state
and pinpoint this field. In particular, the concrete response for 
\texttt{CFGR[SWS]} generated by \sys led the symbolic execution into an infinite loop.
Tracing back, a symbolic value corresponding to \texttt{CFGR[SWS]} was detected.
Then, we read the relevant sentences that
describe \texttt{CFGR[SWS]} in the manual and
found that \texttt{CFGR[SWS]} should follow the value in 
\texttt{CFGR[SW]}. Moreover, 3 is an impossible value. Therefore, we updated this rule as:
\begin{align}
V \quad CFGR[SW] == 0 \rightarrow CFGR[SWS] := 0\\
V \quad CFGR[SW] == 1 \rightarrow CFGR[SWS] := 1\\
V \quad CFGR[SW] == 2 \rightarrow CFGR[SWS] := 2
\end{align}

The condition is any update operations to \texttt{CFGR[SW]} and
the action is to assign the same value to \texttt{CFGR[SWS]}.

\paragraph{Using the Enhanced \ca Rules for Firmware Evaluation.}
With the enhanced rules, especially
those related to clocks, we reran the 66 unit-test samples from \ppim.
\sys achieves a passing rate of 100\%.
We additionally collected 17 new samples that
represent real-world applications or
more complex demo programs with multiple
peripherals and diverse working modes such as DMA.
For example, the PinLock firmware runs on 
a smart lock, which reads a PIN number through a UART interface,
hashes it to compare with a known hash, and sends a signal to unlock
a digital lock if the PIN is correct.
Details of the additional samples are explained in Appendix~\ref{app:addtests}.
\sys successfully emulated all of them,
and neither \ppim nor \uemu can run any of these samples,
with one exception that \uemu succeeded in emulating
\texttt{Shell}.
In Section~\ref{sec:fidelity}, we clearly show that
\sys achieves much better emulation fidelity.

\vspace*{-3mm}
\subsection{Faithfulness of \ca Rules}
\label{sec:faithfulness}

In this section, we evaluate how faithful the \ca rules used in \sys are to
the chip manuals. To this end, a ``ground-truth'' is necessary. We can
manually check all the sentences for a peripheral to construct the
ground-truth; however, this would be a tedious task that depends on individual's
perceptions. To mitigate possible bias, we instead used the working peripheral
models that are already implemented by QEMU developers. These backend models
are written in the C programming language and only implement essential
peripheral logic for correctly emulating firmware. This avoids the
distraction from massive amount of irrelevant sentences in the original
manual. With the C source code for peripheral models, we then manually
checked the program logic and translated them into \ca rules using the same
format as described. In listing~\ref{lst:QEMUUART} of Appendix~\ref
{app:uartrule}, we list a code snippet for the STM32F405 UART backend
emulator. If a C statement can be translated to
a \ca rule, we comment this line with the rule.
For example, at line 5, when QEMU receives any data via the data register,
the SR[RXNE] field is updated. 
Such logic can be translated into the \ca rule \underline{$B \quad \#DR[R] \geq 0 \rightarrow SR[RXNE] := 1$}.
Then, we can compare the \ca rules generated by \sys with the ``ground-truth'' derived from QEMU.

\begin{table}[!t]
\caption{\ca Rule Faithfulness Compared with Rules Used in QEMU Models (STM32F405)}
\vspace*{-3mm}
\label{table:faithfulness}
\centering
\begin{adjustbox}{max width=0.86\columnwidth}
\begin{tabular}{l|rrrr}
\hline
\textbf{Peripheral} & \multicolumn{1}{l}{\textbf{SLOC}} & \multicolumn{1}{c}{\textbf{\begin{tabular}[c]{@{}c@{}}\# Common\\ Rules\end{tabular}}} & \multicolumn{1}{c}{\textbf{\begin{tabular}[c]{@{}c@{}}\# Missing Rules \\ in \sys\end{tabular}}} & \multicolumn{1}{c}{\textbf{\begin{tabular}[c]{@{}c@{}}\# Missing Rule \\ in QEMU\end{tabular}}} \\ \hline
\textbf{UART}        & 145                               & 10                                                                                        & 0                                                                                               & 16                                                                                             \\
\textbf{SPI}         & 123                               & 5                                                                                         & 0                                                                                               & 13                                                                                             \\
\textbf{ADC}         & 200                               & 9                                                                                         & 2                                                                                               & 46                                                                                             \\
\textbf{Timer}       & 221                               & 1                                                                                        & 0                                                                                               & 28                                                                                             \\
\textbf{EXTI}        & 96                                & 23                                                                                        & 0                                                                                               & 29                                                                                              \\
\textbf{SYSCFG}      & 83                                & 7                                                                                         & 0                                                                                               & 1                                                                                              \\ \hline
\end{tabular}
\end{adjustbox}
\vspace*{-6mm}
\end{table}

Following this method, we collected ``ground-truth'' rules
for all QEMU-supported peripherals of STM32F405, including
SYSCFG, UART, ADC, SPI, Timer and EXTI,
from the source code of the latest QEMU release (7.0.0-rc1).
Note \sys supports 20 peripherals for the same chip.
In Table~\ref{table:faithfulness}, we list SLOC (source lines of code)
for QEMU model implementations,
the number of common (and consistent) rules,
the number of missing rules in \sys but present in QEMU,
and the number of missing rules in QEMU but present in \sys.
In counting SLOC, we excluded code that does not contain peripheral logic, such as data structure definitions.

There is a substantial overlap
between the two rule sets,
which represents the most important
peripheral logic.
Our method failed to find two \ca rules for the ADC peripheral that are present in QEMU.
In listing~\ref{lst:QEMUADC}, we 
use the code snippet from QEMU to
show the corresponding logic.
As can be seen, QEMU truncates the received data based on the value
of ADC\_CR1[RES].
However, in the original manual,
it only states that ADC\_CR1[RES] is used to
select the resolution width of the conversion.
No information is given regarding how ADC\_CR1[RES]
impacts truncation precision, which cannot be automatically referred by \sys.
Fortunately, the wrongly truncated data caused by these missing rules
does not influence the emulation capability of \sys.
That is why our diagnostic tool did not find this imperfection.
Existing firmware-guided solutions cannot deal with this imperfection either,
because either branch in listing~\ref{lst:QEMUADC} is acceptable
for emulation.


On the other hand,
there are lots of missing rules in QEMU compared with ours.
The main reason is that QEMU fails to implement many non-standard
peripheral functions.
To ``support'' a peripheral,
QEMU only needs to implement emulation for
the most common logic of this peripheral.
As such, we frequently observed comments similar to
``[**] is not implemented, the registers are included for compability'' in the QEMU source code.
For example,
QEMU currently does not support IRQ or DMA requests for ADC and SPI.
As another example,
based on the manual, the SPI peripheral provides two
main functions, supporting either the SPI protocol or the 
I2S audio protocol. However, QEMU does not support any I2S functions.
We acknowledge that many of the rules
extracted by \sys correspond
to rarely used peripheral functions and 
were never executed in our evaluation.
However, we did obverse 
some critical peripheral logic being
overlooked by QEMU.
For example, based on the STM32F4XX manual,
once the conversion of the selected regular ADC channel is completed,
the \texttt{EOC} (end of conversion) flag in the SR register
should be set.
However, we did not find such logic in QEMU source code.
To confirm this problem, we ran the ADC demo from the official
STM32F4XX SDK package~\cite{stm32f4sdk} with the latest QEMU
and the emulation hung waiting for
\texttt{EOC} to be set.


\vspace*{-2mm}
\subsection{Emulation Fidelity Test}
\label{sec:fidelity}


We used unit-test samples from \ppim to evaluate emulation fidelity,
and compared the fidelity achieved by \sys with those
on \uemu and \ppim,
since these samples are mostly supported by all the
related work.
To get the reference data,
we leveraged the external debug probes (\eg~ST-Link~\cite{stlink} and OpenSDA~\cite{opensda})
to collect traces on real hardware.
Unfortunately, Atmel SAM3X is not equipped with external debugging
capability.
Therefore, we only performed fidelity tests against
firmware samples for NXP FRDM-K64F~\cite{FRDM64F} and STMicroelectronics Nucleo-F103RB~\cite{Nucleof103}.


In the following paragraphs, we first introduce how to collect
traces on real devices and emulators.
Then, we explain a numeric metric 
to quantify the similarity between two execution traces.
It addresses the non-determinism issue
happening during firmware execution (\eg~the same input
on the same hardware can generate different traces).
Lastly, we discuss the results.


\vspace*{-2mm}
\subsubsection{Trace Collection}
\label{sec:trace_collection}


To collect traces on real devices,
we used OpenOCD~\cite{openocd} and an external debugging dongle
to connect the target boards to the remote \textit{gdbserver} provided
by the chip vendors.
Using the debugger, we recorded the program counter
of each instruction execution.
Collecting execution traces of emulators 
was much easier. We directly logged the
starting address of each to-be-executed translation block.
Then, we aligned it with the disassembled firmware code
to reconstruct the execution trace 
following the same format as that on the real device.

\vspace*{-2mm}
\subsubsection{Metrics}

To quantitatively measure emulation fidelity, 
we propose a new metric to indicate the similarity between
execution traces.
In particular, we use the execution trace on the real device
as the reference.
By representing the traces as a sequence of addresses,
we can use traditional string distance algorithms
such as Levenshtein distance (a.k.a edit distance)~\cite{WiKiED}.
However, due to the non-determinism of firmware execution (\eg~interrupt timing),
directly using edit distance cannot reliably measure the true similarity.
For example, with the same input and firmware,
two executions on real device could yield very different traces,
but both of them achieve 100\% fidelity.

To address this issue,
we divided each trace into three 
parts based on their functions: 
1) The \textbf{initialization trace} covers the initialization functions like peripheral configurations.
It ends at the start of the main function.
2) The \textbf{main loop trace} includes the main firmware
business.
In the unit-tests, it corresponds to the data transmission logic.
We modified the unit-tests so that the main loop was executed five times. 
3) The \textbf{interrupt trace} records the instructions
executed in the interrupt context.
We separate it from others to eliminate non-determinism.


For each part of trace, we use edit distance to measure the similarity.
In computational linguistics and computer science, edit distance
is used to quantify how dissimilar two strings are,
specifically, what is the minimum operations
 (\ie~deletion, insertion and substitution) needed to transfer
one to another.
Similarly,
we can use this method to measure the distance from
traces on an emulator to traces on the real device.
To be specific,
the whole trace is presented as a sequence of addresses,
which correspond to the starts of each basic block.
Compared with the trace of real device,
a deletion operation means the emulator mistakenly
takes an additional basic block;
an insertion operation means the emulator misses a basic block execution;
a substitution operation means the emulator executes a different basic block.
The fewer operations are needed,
the more similar are the two traces.
In the original algorithm, deletion, insertion and substitution
weigh equally. However, in our case,
deleting or inserting repeated sequences are considered less important.
Indeed, missing some code execution or executing some wrong code
usually have worse effects on firmware analysis
compared with repeating previous traces.
For example, firmware usually runs in a polling mode
that waits for a certain value of the status register.
Whether it runs 100 loops or 1,000 loops does not
effectively influence firmware analysis.
Therefore, we assign the weight of repeated deletion or insertion
operations to 1 while the others to 2.


The trace distance between the emulator and real device is calculated
by $ D_{Emulator} = D_{init}+D_{main}+D_{irq}$,
where $D_{init}$, $D_{main}$, and $D_{irq}$ are the modified edit distances
for the three parts of trace mentioned before.
Finally, we normalize the measured distances into a number ranging from 0 to 1
to represent the fidelity score,
with 1 being the most similar (\ie~short distance).
Concretely, the fidelity score of an emulator can be calculated by
$1 -  \min(\frac{D_{Emulator}}{D_{QEMU}}, 1)$, where $D_{QEMU}$ is the edit distance
measured on unmodified QEMU.
Here, $D_{QEMU}$ represents the worst case emulation result because
no peripheral emulation is provided.
As can be seen, if $D_{Emulator}$ is low enough, the fidelity score will be 1.
If $D_{Emulator}$ is equal to or higher
than $D_{QEMU}$, the fidelity score will be 0.
Otherwise, the fidelity score ranges from 0 to 1.







\vspace*{-3mm}
\subsubsection{Results}
\label{sec:fidelityres}
In Table~\ref{tab:score}, we show a break-down
of the fidelity scores for the three parts of a trace with I2C, UART,GPIO and Timer unit test samples.
Under the proposed fidelity metric,
\sys is the only one that has perfect scores
among the test samples.

The most significant difference occurs in the interrupt trace,
where both \ppim and \uemu perform very badly.
After manual analysis,
we found that \ppim mis-categorizes many control or status registers as data registers, which triggers unexpected emulation results.
For example, when the transmit enable field (\texttt{CR[TE]}) is overwritten
by external data, 
it could prevent firmware from executing the data transmission function.
\uemu, on the other hand,
suffers from low fidelity issues due to its unsound invalid state heuristics.
For example, we found it frequently invoked infeasible paths
which do not actually cause a hang or crash. 
Finally, neither \ppim nor \uemu is aware of the
interrupt timing. Therefore, the interrupt handler typically
takes irrelevant paths.

\vspace*{-4mm}
\subsection{Fuzz Testing}
\label{sec:fuzz}


In this section, we evaluate \sys regarding fuzzing.
The tested firmware includes 10 samples used in \ppim~\cite{p2imgit} and
2 complex samples used in Pretender~\cite{pretednergit}.
We also added 4 new samples 
that perform network communication over TCP and UDP,
using LwIP over Ethernet as network layer protocol.
They were developed based on the demo code for the STM32F429 chip.
We integrated modified AFL as the fuzzing
engine. 
Due to the randomness of fuzzing,
we conducted 5 trials of 24-hour fuzzing
for each test sample
to keep the experiment in line with community guidelines~\cite{klees2018evaluating}.
We used the same random value as the initial
seed for each test target (\ppim, \uemu, \sys).
Where applicable,
we used the default configurations published alongside the
released tools.



\begin{table*}
\setlength\abovedisplayskip{0pt}
\setlength\belowdisplayskip{0pt}
\caption{Fuzzing and Compliance Check Results (- indicates
the emulator cannot support fuzzing that firmware)}
\vspace{-3mm}
\centering
\begin{adjustbox}{max width=0.95\textwidth}

\setlength\abovedisplayskip{0pt}
\setlength\belowdisplayskip{0pt}
\begin{tabular}{l|l|rrrrr|crrrrrr|c}
\hline
                  &                                   & \multicolumn{3}{c}{\textbf{\begin{tabular}[c]{@{}c@{}}Median\\ BB Coverage\end{tabular}}}    & \multicolumn{2}{c|}{\textbf{\textit{p}-value}}                  & \multicolumn{4}{c}{\textbf{\#Crashes/\#Hangs}}                                                                                     & \multicolumn{3}{c|}{\textbf{\#False Crashes/Hangs}}                                           & \textbf{\begin{tabular}[c]{@{}c@{}}Compliance\\ Violation\end{tabular}} \\
\textbf{Firmware} & \multicolumn{1}{c|}{\textbf{MCU}} & \multicolumn{1}{c}{\textbf{\sys}} & \multicolumn{1}{c}{\ppim} & \multicolumn{1}{c|}{\uemu}   & \multicolumn{1}{c}{\textit{to} \ppim} & \multicolumn{1}{c|}{\textit{to} \uemu} & \multicolumn{1}{c|}{\textbf{Unique}} & \multicolumn{1}{c}{\textbf{\sys}} & \multicolumn{1}{c}{\ppim} & \multicolumn{1}{c|}{\uemu} & \multicolumn{1}{c}{\textbf{\sys}} & \multicolumn{1}{c}{\ppim} & \multicolumn{1}{c|}{\uemu} & \textbf{\sys}                                                           \\ \hline
CNC               & STM32F429                         & \textbf{30.90\%}                  & 34.31\%                   & \multicolumn{1}{r|}{28.52\%} & 0.21                      & 0.30                       & \multicolumn{1}{c|}{\textbf{0/0}}  & \textbf{0/0}                      & 1/3                       & \multicolumn{1}{r|}{0/0}   & \textbf{0/0}                      & 1/3                       & 0/0                        & \textbf{None}                                                           \\
Console           & K64F                              & \textbf{30.20\%}                  & 33.12\%                   & \multicolumn{1}{r|}{29.31\%} & <0.01                     & <0.01                      & \multicolumn{1}{c|}{\textbf{0/0}}  & \textbf{0/0}                      & 1/3                       & \multicolumn{1}{r|}{0/0}   & \textbf{0/0}                      & 1/3                       & 0/0                        & \textbf{None}                                                           \\
Drone             & STM32F103                         & \textbf{57.59\%}                  & 46.67\%                   & \multicolumn{1}{r|}{53.48\%} & <0.01                     & <0.01                      & \multicolumn{1}{c|}{\textbf{0/0}}  & \textbf{0/0}                      & 1/7                       & \multicolumn{1}{r|}{0/0}   & \textbf{0/0}                      & 1/7                       & 0/0                        & \textbf{None}                                                           \\
Gateway           & STM32F103                         & \textbf{36.52\%}                  & 36.20\%                   & \multicolumn{1}{r|}{36.66\%} & 0.92                      & 0.23                       & \multicolumn{1}{c|}{\textbf{1/0}}  & \textbf{3/0}                      & 136/254                   & \multicolumn{1}{r|}{3/0}   & \textbf{0/0}                      & 7/254                     & 0/0                        & \textbf{None}                                                           \\
HeatPress         & SAM3X8E                           & \textbf{28.13\%}                  & 29.32\%                   & \multicolumn{1}{r|}{26.88\%} & <0.01                     & <0.01                      & \multicolumn{1}{c|}{\textbf{2/0}}  & \textbf{9/0}                      & 161/48                    & \multicolumn{1}{r|}{4/0}   & \textbf{0/0}                      & 8/48                      & 0/0                        & \textbf{None}                                                           \\
PLC               & STM32F429                         & \textbf{23.13\%}                  & 22.14\%                   & \multicolumn{1}{r|}{19.84\%} & <0.01                     & <0.01                      & \multicolumn{1}{c|}{\textbf{5/0}}  & \textbf{13/0}                     & 85/27                     & \multicolumn{1}{r|}{65/0}  & \textbf{0/0}                      & 5/27                      & 0/0                        & \textbf{None}                                                           \\
Reflow\_Oven      & STM32F103                         & \textbf{35.67\%}                  & 27.97\%                   & \multicolumn{1}{r|}{29.84\%} & <0.01                     & <0.01                      & \multicolumn{1}{c|}{\textbf{0/0}}  & \textbf{0/0}                      & 1/0                       & \multicolumn{1}{r|}{0/0}   & \textbf{0/0}                      & 1/0                       & 0/0                        & \textbf{None}                                                           \\
Robot             & STM32F103                         & \textbf{39.99\%}                  & 36.90\%                   & \multicolumn{1}{r|}{35.91\%} & <0.01                     & <0.01                      & \multicolumn{1}{c|}{\textbf{0/0}}  & \textbf{0/0}                      & 1/0                       & \multicolumn{1}{r|}{0/0}   & \textbf{0/0}                      & 1/0                       & 0/0                        & \textbf{R2(B)}                                                          \\
Soldering\_Iron   & STM32F103                         & \textbf{48.92\%}                  & 37.60\%                   & \multicolumn{1}{r|}{35.03\%} & <0.01                     & <0.01                      & \multicolumn{1}{c|}{\textbf{0/0}}  & \textbf{0/0}                      & 1/5                       & \multicolumn{1}{r|}{38/7}  & \textbf{0/0}                      & 1/5                       & 38/7                       & \textbf{R2(A)}                                                          \\
Steering\_Control & SAM3X8E                           & \textbf{26.65\%}                  & 27.07\%                   & \multicolumn{1}{r|}{27.57\%} & <0.01                     & <0.01                      & \multicolumn{1}{c|}{\textbf{0/0}}  & \textbf{0/0}                      & 1/6                       & \multicolumn{1}{r|}{3/0}   & \textbf{0/0}                      & 1/6                       & 2/0                        & \textbf{None}                                                           \\
RF\_Door\_Lock    & STM32L152                         & \textbf{21.20\%}                  & -                         & \multicolumn{1}{r|}{20.86\%} & -                         & <0.01                      & \multicolumn{1}{c|}{\textbf{2/0}}  & \textbf{13/0}                     & -                         & \multicolumn{1}{r|}{119/0} & 0/0                               & -                         & 0/0                        & \textbf{R2(A)}                                                          \\
Thermostat        & STM32L152                         & \textbf{23.59\%}                  & -                         & \multicolumn{1}{r|}{22.54\%} & -                         & <0.01                      & \multicolumn{1}{c|}{\textbf{1/0}}  & \textbf{7/0}                      & -                         & \multicolumn{1}{r|}{97/0}  & 0/0                               & -                         & 0/0                        & \textbf{R2(A)}                                                          \\ \hline
LwIP\_TCP\_Client & STM32F429                         & \textbf{29.45\%}                  & -                         & \multicolumn{1}{r|}{-}       & -                         & -                          & \multicolumn{1}{c|}{\textbf{0/0}}  & \textbf{0/0}                      & -                         & \multicolumn{1}{r|}{-}     & \textbf{0/0}                      & -                         & -                          & \textbf{None}                                                           \\
LwIP\_TCP\_Server & STM32F429                         & \textbf{29.40\%}                  & -                         & \multicolumn{1}{r|}{-}       & -                         & -                          & \multicolumn{1}{c|}{\textbf{0/0}}  & \textbf{0/0}                      & -                         & \multicolumn{1}{r|}{-}     & \textbf{0/0}                      & -                         & -                          & \textbf{None}                                                           \\
LwIP\_UDP\_Client & STM32F429                         & \textbf{29.75\%}                  & -                         & \multicolumn{1}{r|}{-}       & -                         & -                          & \multicolumn{1}{c|}{\textbf{0/0}}  & \textbf{0/0}                      & -                         & \multicolumn{1}{r|}{-}     & \textbf{0/0}                      & -                         & -                          & \textbf{None}                                                           \\
LwIP\_UDP\_Server & STM32F429                         & \textbf{30.36\%}                  & -                         & \multicolumn{1}{r|}{-}       & -                         & -                          & \multicolumn{1}{c|}{\textbf{0/0}}  & \textbf{0/0}                      & -                         & \multicolumn{1}{r|}{-}     & \textbf{0/0}                      & -                         & -                          & \textbf{None}                                                           \\ \hline
\end{tabular}
\end{adjustbox}
\label{tab:fuzz}
\flushleft
\vspace{-5mm}
\end{table*}



\paragraph{Coverage Comparison.}
The results of the fuzzing experiment is shown
in Table~\ref{tab:fuzz}.
We list the median basic block (BB)
coverage in the 5 trials, along with the p-value of the experiments.
We observe obvious improvement in code coverage over \ppim and \uemu
in \texttt{Soldering\_Iron} and four TCP/UDP samples.
The reason is that neither \ppim nor \uemu supports
DMA and thus Ethernet.
Although \uemu is able to go through the initialization of the Ethernet samples, 
it failed to cover the main firmware logic
which depends on DMA for data transmission.
For the firmware that can be tested,
\sys yields 5.1\% and 6.7\% more code coverage on average 
than \ppim and \uemu, respectively.
For each sample, we also list the p-value of the tests.
They are measured using two-side Mann-Whitney U test
for \sys vs. \ppim, and \sys vs. \uemu.

The coverage improvement seems insignificant.
However, we note that \sys achieves better path fidelity
and so does not explore many error handling functions.
For example, some peripherals like I2C use 
two separated interrupts to deal with normal and error signals.
The error interrupt only occurs on hardware fault.
However, \ppim and \uemu still trigger these error interrupts periodically,
which should never happen without incurring physical failure.
In addition,
\sys only fuzzes real I/O interfaces,
not the status/control registers which existing work also fuzzes.
Because of these reasons, \sys achieves less code coverage on some firmware (\eg~Steering\_Control).
We argue this is a good indicator of higher emulation fidelity.

\paragraph{Crashes/Hangs.}
In Table~\ref{tab:fuzz}, we list
the number of crashes and hangs reported by AFL
in the column \textbf{\#Crashes/\#Hangs}.
To verify the results, for each
reported crash/hang, we replayed the trigger test-case to
the corresponding emulator,
and collected execution traces.
If two execution traces are the same, we consider it as duplicate.
If we did not observe the crash/hang report during replay,
we consider it a false positive.
We list the unique crashes/hangs in the column with 
the header \textbf{Unique},
and the number of false crashes/hangs in the column with the header \textbf{\#False Crashes/Hangs}.

\sys successfully reproduced all the bugs mentioned in previous work,
but did not report any false crashes or hangs.
In fact, many crash reports raised in
existing work were duplicated or false positives.
The reasons are two-fold.
First, existing work randomly delivers interrupts,
which causes significant changes in the edge coverage
of two executions that suffer from the same crash.
AFL would thus mistakenly recognize them as two different crashes.
Second, as indicated before, due to inaccurate emulation in \ppim and \uemu,
there are many infeasible paths being explored,
leading to considerable occurrences of false positives.

\vspace*{-5mm}
\subsection{Compliance Check Test}
\label{sec:comp}

With semantic information extracted from the specification,
and benefited from the high fidelity emulation,
\sys makes another kind of security analysis,
namely behavior compliance check, possible.
In a compliance check, we 
check whether the implementation of peripheral 
drivers follows the descriptions from the chip manuals.
Specifically, we collect the peripheral access history
and match it with the expected access constraints learned by NLP.
In our prototype, we implemented a dynamic compliance check
module that checks two rules.

\paragraph{R1: Peripheral State Verification.}
The firmware should only access certain registers
after checking the status of another register.
This simple rule applies to many I/O operations.
For example, the firmware first checks whether
the hardware is ready by polling the state from
a status register.
Only if a particular value is returned
would it continue to access the data register.
R1 is violated if the firmware directly
accesses the data register.


\paragraph{R2: Interrupt Activation Consistency.}
To enable an interrupt,
the firmware not only needs to activate the configuration
local to the peripheral, but also to enable the
corresponding interrupt source by setting the
Interrupt Set-Enable Registers (\texttt{ISERx}) in the
global interrupt manager, namely NVIC.
A violation happens when:
\textbf{R2(A):} the firmware enables the interrupt in NVIC,
but not in the local peripheral controller.
\textbf{R2(B):} the firmware enables the local peripheral controller, 
but not in NVIC.

\vspace{-3mm}
\subsubsection{Experiment Results.}
We performed compliance check
against the same set of firmware samples used in
the previous section,
and the unit-test samples of \ppim.
The results are as shown in the last column of Table~\ref{tab:fuzz} and in the Table~\ref{tab:comp} 
in Appendix.

\paragraph{R1 Violation.}
We observed several R1 violations in the SAM3X HAL driver code.
We later confirmed the root cause to be
race conditions of peripheral access.
For example, the drive code for
UART of SAM3X MCU checks the \texttt{TXRDY} field before data transmission.
However, it also allows a UART interrupt to happen
between \texttt{TXRDY} verification and data transmission,
during which the interrupt handler conducts an independent
data transmission.
When the interrupt returns,
the previous \texttt{TXRDY} verification becomes
invalided and the following data transmission would fail.
We found similar issues with the SPI unit-test sample on STMF103 MCU.

\paragraph{R2 Violation.}
Chip vendors often provide a HAL library for developers.
It abstracts away hardware details and provides a unified
API to access peripheral functions.
However, 
not all HALs support the NVIC register configuration.
This is because the IRQ number assignment is chip-specific,
and it is developer's responsible to configure NVIC.
This assumption gives rise to the violation of R2.
In particular, the developer may forget configuring NVIC.
For example, we found the \texttt{Robot} firmware invokes the HAL 
function \texttt{HAL\_TIM\_Base\_Start\_IT()} to set the \texttt{TIM2\_DIER} register to enable the \texttt{Timer2} interrupt.
However, it does not enable the corresponding interrupt number (28)
in the NVIC registers.
As a result, the interrupt can never be delivered.
This happens to the original K64F Timer unit test sample too.
After manual verification,
we found the test code enables a non-existing
interrupt channel for the K64F Timer,
which has been confirmed and corrected by \ppim authors\footnote{\url{https://github.com/RiS3-Lab/p2im-unit_tests/commit/cdfd9bbc72e1cc87a1f5d3905804ae1f91539beb}}.
We observed similar violations happened
to \texttt{Soldering\_Iron},
\texttt{RF\_Door\_Lock}, and \texttt{Thermostat}.

\vspace{-2mm}
\section{Limitations and Discussion}


\paragraph{NLP Limitations.} \label{sec::nlplimitation}
The state-of-the-art NLP tools have limitations in handling references across different sentences and complex sentences with underlying or nested conditions.
These limitations cause faulty rules as discussed in Section~\ref{sec:enhance}.
However, unlike other applications, 
\sys suffers less from these limitation because 
1) actions are often explicitly expressed as executing several simple register assignments and no cross-section reference is used; 
2) named entities are formal expressions in chip manuals, alleviating co-reference problems (\eg~we use approximate string matching
to unify named entities); 
3) complex conditions often depend on certain \hw
with implicit semantics, which can often be modelled
by a default behavior (\eg~the always ``set'' heuristic
we adopted in the F103 ADC example in Section~\ref{sec:modelSynthesizing}).
We also plan to leverage new advances in NLP techniques 
to improve \sys in the future.
For example, \sys can greatly benefit from the improvement on
co-reference resolution in deciding the conditions
of hardware triggered signals.

\paragraph{Manual Efforts.} \label{sec::manualeffort}
There are three major sources of manual efforts.
First, for each firmware, the developers need to identify the
underlying MUC chip and configure the memory 
mapping information (\eg~the ranges of flash, RAM, MMIO) for the emulator.
Second, the developers need to manually copy the required
sections in the PDF-format chip manuals and input them to the NLP engine.
Currently, register memory map, field description, interrupt vector assignment table and DMA channel assignment table, or
equivalent sections are needed.
Third, diagnosing faulty \ca rules requires two-fold manual efforts.
1) Test firmware and test-cases need to be prepared;
2) When an invalid state is detected, 
the developers need to read and comprehend
the specification to fine-tune the faulty \ca rules.
The latter cannot be avoided since there are
many domain terminologies which cannot be understandable by NLP.
For example, the F1XX manual mentions
that I2C can enter different modes depending on 
the LSB of the address byte. 
The knowledge that LSB refers to ``the low-order bit 
of the transmit buffer'' must be provided by experts.

To evaluate the manual effort needed
in \ca rule diagnosis,
we invited three embedded system developers with
basic domain knowledge on MCU.
After reading through the relevant chapters of the manual,
two hours were needed to fix the problematic \ca rules
for F103 I2C on average.
In comparison, we estimate that 
it would take more than a week for an expert 
to write an I2C backend emulator for QEMU.

\vspace*{-3mm}
\section{Related Work}

\subsection{Firmware Emulation}
\vspace*{-1mm}
Relying on real hardware for dynamic analysis
incurs many problems~\cite{muench2018avatar2, koscher2015surrogates, kammerstetter2016embedded, kammerstetter2014prospect}, 
such as poor performance and low scalability.
Emulation is an effective way to address these issues.
The key challenge of emulating firmware is
how to properly model the peripheral behaviors so that
the emulated execution can be similar to that on real hardware.


High-level emulation solutions~\cite{clements2020halucinator,shoshitaishvili2015firmalice, chen2016towards, costin2016automated,wen2021from}
avoid emulating code related to peripheral by hooking
into high-level libraries (\eg~hardware abstraction layer) and implementing equivalent logic on the 
native machine.
High-level emulation achieves good fidelity as \sys does.
However, these approaches completely skip the peripheral logic in firmware,
and therefore cannot find problems with peripheral drivers.
Furthermore, 
developers do not always use a high-level abstraction library
for performance consideration.


Firmware-guided solutions~\cite{fengp2020p2im,chen2016towards,spensky2021conware,cao2020device, zhou2021automatic,johnson2021jetset}
run the whole target firmware in the emulator.
Based on the strategy,
they can be further classified into three types:
access-pattern-based, symbolic-execution-based,
and learning-based.
By observing the peripheral access pattern,
\ppim infers register types (\ie~\texttt{CR}, \texttt{SR}, \texttt{C\&SR} and \texttt{DR}),
then it uses heuristics to generates responses based on
the register type information.
However, as mentioned before, it suffers from the register
mis-categorization problem. 
Besides, when heuristic is unavailable, \ppim blindly searches
for appropriate responses with limited search spaces.
Symbolic-execution-based solutions~\cite{cao2020device,zhou2021automatic,johnson2021jetset} address
the aforementioned problems by reasoning about how
responses from peripheral can influence firmware execution.
A key limitation of these solutions is that they
rely on heuristics to decide the path to take.
Indeed, firmware does not contain enough information
to guide the emulation.
PRETENDER~\cite{gustafson2019toward} and Conware~\cite{spensky2021conware} create peripheral models by learning from the real interactions 
between the hardware and firmware. 
As such, they require a hardware dependent recording phase,
reducing the scalability.
Compared with firmware-guided solutions, our approach 
is guided by peripheral specifications.
It achieves higher emulation fidelity without requiring real hardware.


\vspace{-3mm}
\subsection{Rule Extraction from Specifications via NLP Technology}
\label{sec::nlpwork}


Extracting rules from specifications to solve security problems is not new.
SmartAuth~\cite{tian2017smartauth} learns the policy correlations (entity, context and action, condition) from IoT app descriptions, and verifies whether the code implementation of IoT apps follows the policy. 
ARE~\cite{feng2018acquisitional} automatically discovers IoT devices by generating (device detection) rules using application-layer data and 
product descriptions. 
iRuler~\cite{wang2019charting} uses NLP techniques to infer trigger-action information flows and discover inter-rule vulnerabilities within IoT deployments.  
NLP techniques are also used to automatically collect and analyze IoT security reports to extract vulnerability-specific features~\cite{feng2019understanding}. 
Bookworm Game~\cite{bookworm2021} uses NLP techniques 
to identify risky operations 
from LTE (Long-Term Evolution) documentation in telecommunication.
Then it extracts information to construct testcases
that can satisfy the conditions to trigger risky operations.
Access control policy~\cite{xiao2012automated} and access control rules~\cite{slankas2014relation} can also be
extracted from use case descriptions.

Our work automatically constructs peripheral models for 
security-analysis-driven firmware emulation.
This is a different NLP application from existing work.
This specific problem introduces a few unique challenges.
First, a \ca rule can be triggered in different ways.
We must categorize the extracted rules so that they
can be checked only when needed.
Second, to support firmware emulation, the semantics of the extracted rules must be formalized into concrete values and Boolean operators so that the peripheral states can be programmatically
maintained.
Third, in order to handle hardware-generated signals and interrupt-related actions, we must incorporate MCU-specific domain-knowledge into NLP.
Most importantly,
existing work does not need to comprehensively extract relevant rules, but \sys must ensure that the extracted rules are adequately complete. 
For example, in Bookworm Game~\cite{bookworm2021}, the more context information it can extract, the more risky operations it could identify. 
Even if the extracted information items are only a subset,
their approach can still outperform solutions not 
guided by NLP techniques. 
In contrast, if \sys does not achieve adequate 
completeness of the relevant \ca rules (including the chained rules), 
firmware emulation will very likely fail. 
This key difference motivates the unique rule diagnosis 
mechanism of \sys.   

\vspace*{-1.5mm}
\section{Conclusion}

Instead of proposing yet another firmware-guided emulation solution, 
in this work we propose the first specification-based firmware emulation solution. 
The new approach leverages   
NLP techniques to translate peripheral behaviors (specified) in human 
language (documented in chip manuals) into a set of 
structured condition-action rules. 
By properly executing and chaining these rules at runtime,
we can dynamically synthesize a peripheral model for each peripheral 
accessed during firmware execution. 
With the help of machine-aided rule diagnosis,
our evaluation confirmed 
that our prototype achieves 100\% emulation fidelity 
compared with real devices
under the proposed fidelity measurement.
In comparison, the fidelity achieved by existing work
varies between 73\% and 86\%. 
With better fidelity, our solution improves fuzzing efficiency: no false crashes and hangs were observed during fuzz testing.  
With much higher emulation accuracy,
we also designed a new dynamic analysis task to perform 
driver code compliance checks against the specification. 
We found some non-compliance which
we later confirmed to be bugs caused by race condition.


\vspace*{-1mm}
\bibliographystyle{ACM-Reference-Format}
\bibliography{bibs/main}

\appendix
\section{The prevalence of chip specification among top MCU chip vendors}
\label{app:mcus}

Table~\ref{tab:mcus} shows the result of our survey about
the availability of chip specifications among top MCU chip vendors.
We also include a link to an example manual for each vendor.

\begin{table}[!t]   
    \caption{The prevalence of MCU specifications}
\label{tab:mcus}
\centering
\begin{adjustbox}{max width=\columnwidth}
\begin{tabular}{l|llll}
\hline
\textbf{Chip vendor}           &\multicolumn{1}{c}{\begin{tabular}[c]{@{}c@{}}Have public\\  manuals\end{tabular}} & \multicolumn{1}{c}{\begin{tabular}[c]{@{}c@{}}Require\\ registration\end{tabular}} & \begin{tabular}[c]{@{}l@{}}Have the needed \\ information\end{tabular} & \begin{tabular}[c]{@{}l@{}}Example\\ Manual\end{tabular} \\ \hline
\textbf{Texas Instruments}     & Yes                            & No                            & Yes                                  &  \cite{Texas}                    \\
\textbf{Maxim Integrated}      & Yes                            & No                            & Yes                                  &    \cite{Maxim}                  \\
\textbf{NXP}                   & Yes                            & Yes                           & Yes                                  &    \cite{nxp}                  \\
\textbf{Microchip}             & Yes                            & No                            & Yes                                  &    \cite{microchip}                  \\
\textbf{Infineon Technologies} & Yes                            & No                            & Yes                                  &    \cite{infineon}                  \\
\textbf{STMicroelectronics}    & Yes                            & No                            & Yes                                  &    \cite{stm32}                  \\
\textbf{Analog Device}         & Yes                            & No                            & Yes                                  &    \cite{Analog}                  \\
\textbf{Renesas}               & Yes                            & No                            & Yes                                  &    \cite{Renesas}                  \\
\textbf{Silicon Labs}          & Yes                            & No                            & Yes                                  &    \cite{Silicon}                  \\
\textbf{Toshiba}               & Yes                            & No                            & Yes                                  &    \cite{Toshiba}                  \\
\textbf{Nuvoton Technology}    & Yes                            & No                            & Yes                                  &    \cite{Nuvoton}                  \\
\textbf{ZiLOG}                 & Yes                            & No                            & Yes                                  &    \cite{ZiLOG}                  \\
\textbf{ON Semiconductor}      & Yes                            & No                            & Yes                                  &    \cite{Semiconductor}                  \\
\textbf{Nordic}                & Yes                            & No                            & Yes                                  &    \cite{nordic}                  \\ \hline
\end{tabular}
\end{adjustbox}
\end{table}

\vspace*{-2mm}
\section{Formal Representation of \ca Rules}
\label{app:carule}
We formalize the conditions and actions below.

Type 1 Condition:
\begin{equation}\label{equ:cond3}
    B\ \#D[R/T] ==/\geq / \leq  value/Reg[Field]
\end{equation}
\begin{equation}\label{equ:cond4}
    O *
\end{equation}

Type 2 Condition:
\begin{equation}\label{equ:cond2}
    R/W\ Reg[Field]
\end{equation}

Type 3 Condition:
\begin{equation}\label{equ:cond1}
    V\ Reg[Field] ==/\geq / \leq  value/Reg[Field]
\end{equation}

Type 1 Action:
\begin{equation}\label{equ:act1}
    Reg[Field] := value/Reg[Field]
\end{equation}

Type 2 Action:
\begin{equation}\label{equ:act2}
    IRQ[InterruptSource] := Disable/Enable/Pending
\end{equation}

Type 3 Action:
\begin{equation}\label{equ:act3}
    DMA[InterruptSource] := Disable/Enable/Pending
\end{equation}

\section{Extracted \ca Rules for K64F UART0}
\label{app:uartrule}
In Listing~\ref{tab:k64uart1carule},
we first enumerate the field name, its address, and bits. Here \texttt{*} in \texttt{Bits} Column represents all bits in that registers.
Then we list the interrupt sources and the DMA sources.
Lastly,  we show a sample of extracted \ca rules.
\begin{lstlisting}[style=CStyle,numbers=none,label={tab:k64uart1carule},xleftmargin=1em,framexleftmargin=1em,frame=shadowbox,caption={A Sample of \ca Rules for K64F UART0}]
Field Name              Address                   Bits
C2[ILIE]                0x4006a003                   4
C2[RIE]                 0x4006a003                   5
C2[TCIE]                0x4006a003                   6
C2[TIE]                 0x4006a003                   7
S1[OR]                  0x4006a004                   3
S1[IDLE]                0x4006a004                   4
S1[RDRF]                0x4006a004                   5
S1[TC]                  0x4006a004                   6
S1[TDRE]                0x4006a004                   7
S2[LBKDIF]              0x4006a005                   7
C3[ORIE]                0x4006a006                   3
D[T]                    0x4006a007                   *
D[R]                    0x4006a007                   *
C5[ILDMAS]              0x4006a00b                   4
C5[RDMAS]               0x4006a00b                   5
C5[TCDMAS]              0x4006a00b                   6
C5[TDMAS]               0x4006a00b                   7
CFIFO[RXUFE]            0x4006a011                   0
CFIFO[TXOFE]            0x4006a011                   1
SFIFO[RXUF]             0x4006a012                   0
SFIFO[TXOF]             0x4006a012                   1
TWFIFO[TXWATER]         0x4006a013                   *
TCFIFO[TXCOUNT]         0x4006a014                   *
RWFIFO[RXWATER]         0x4006a015                   *
RCFIFO[RXCOUNT]         0x4006a016                   *
...
------------------------------------------------------
Interrupt Source                            IRQ Number
TDRE                                                31
TC                                                  31
IDLE                                                31
RDRF                                                31
RXUF                                                31
TXOF                                                31
...
------------------------------------------------------
DMA Source                                  IRQ Number
Channel 0 transfer complete                          0
Channel 1 transfer complete                          1
...
Error interrupt Channels 0-15                       16
------------------------------------------------------
Rules
B #D[T] > TWFIFO[TXWATER]  -> S1[TDRE] := 0 
B #D[T] <= TWFIFO[TXWATER]  -> S1[TDRE] := 1 
B #D[R] >= RWFIFO[RXWATER]  -> S1[RDRF] := 1 
B #D[R] < RWFIFO[RXWATER] -> S1[RDRF] := 0 
V S2[LBKDIF] == 1 -> S2[LBKDIF] := 0 
V SFIFO[TXOF] == 1 -> SFIFO[TXOF] := 0 
V SFIFO[RXUF] == 1 -> SFIFO[RXUF] := 0 
W C2[TIE] == *-> C2[TIE] = * 
W C5[TDMAS] == *-> C5[TDMAS] = * 
R D[R] == * -> S1[IDLE] := 0 
R D[R] == * -> S1[OR] := 0 
O * -> RCFIFO[RXCOUNT] == #D[R] 
O * -> TCFIFO[TXCOUNT] == #D[T] 
V S1[TDRE] == 1 & V C2[TIE] == 1 & V C5[TDMAS] == 1 -> DMA[TDRE] := Ready 
V S1[TDRE] == 1 & V C2[TIE] == 1 & V C5[TDMAS] == 0 -> IRQ[TDRE] := Ready 
V S1[TC] == 1 & V C2[TCIE] == 1 & V C5[TCDMAS] == 1 -> DMA[TC] := Ready 
V S1[TC] == 1 & V C2[TCIE] == 1 & V C5[TCDMAS] == 0 -> IRQ[TC] := Ready 
V S1[IDLE] == 1 & V C2[ILIE] == 1 & V C5[ILDMAS] == 1 -> DMA[IDLE] := Ready 
V S1[IDLE] == 1 & V C2[ILIE] == 1 & V C5[ILDMAS] == 0 -> IRQ[IDLE] := Ready 
V S1[RDRF] == 1 & V C2[RIE] == 1 & V C5[RDMAS] == 1 -> DMA[RDRF] := Ready 
V S1[RDRF] == 1 & V C2[RIE] == 1 & V C5[RDMAS] == 0 -> IRQ[RDRF] := Ready 
V S1[OR] == 1 & V C3[ORIE] == 1 -> IRQ[OR] := Ready 
V SFIFO[RXUF] == 1 & V CFIFO[RXUFE] == 1 -> IRQ[RXUF] := Ready 
V SFIFO[TXOF] == 1 & V CFIFO[TXOFE] == 1 -> IRQ[TXOF] := Ready 
...
\end{lstlisting}

\section{Compliance Check Results of \ppim Unit-Test Samples}
Table~\ref{tab:comp} shows the compliance check results of \ppim unit-test samples.
\begin{table}[!ht]
\caption{Compliance Check Results of \ppim Unit-Test Samples}
\label{tab:comp}
\centering
\begin{adjustbox}{max width=0.6\columnwidth}
\begin{tabular}{l|l|c|r}
\hline
\multicolumn{1}{c|}{\textbf{Peri.}} & \multicolumn{1}{c|}{\textbf{MCU}} & \textbf{OS} & \multicolumn{1}{c}{\textbf{\begin{tabular}[c]{@{}c@{}}Compliance\\ Violation\end{tabular}}} \\ \hline
\multirow{3}{*}{UART}               & K64F                              & RIOT        & R1                                                                                          \\
                                    & SAM3X                             & RIOT        & R1                                                                                          \\
                                    & SAM3X                             & ARDUINO     & R1                                                                                          \\\hdashline[1pt/5pt]
I2C                                 & SAM3X                             & ARDUINO     & R2(B)                                                                                       \\\hdashline[1pt/5pt]
SPI                                 & F103                              & ARDUINO     & R1                                                                                          \\\hdashline[1pt/5pt]
GPIO                                & SAM3X                             & ARDUINO     & R2(B)                                                                                       \\\hdashline[1pt/5pt]
TIMER                               & K64F                              & RIOT        & R2(A)                                                                                       \\ \hline
\end{tabular}
\end{adjustbox}
\end{table}

\section{Statistics of \ca Rules }
Table~\ref{tab:nlp} shows the statistics 
of \ca rules for the evaluated peripherals.


\begin{table}[!ht]   
    \caption{\ca Rule Statistics}
\label{tab:nlp}
\vspace*{-3mm}
\centering
\begin{adjustbox}{max width=\columnwidth}
\begin{tabular}{l|lrr|rrr|rrr|r}
\hline
      &      &       &  \textbf{\#Reg.}            & \multicolumn{3}{c|}{\textbf{\#Conditions}} & \multicolumn{3}{c|}{\textbf{\#Actions}} & \textbf{Total}    \\

\textbf{Peri.} & \textbf{MCU}  & \textbf{\#Sent.} & \textbf{(Fields)} & \textbf{C1}    & \textbf{C2}    & \textbf{C3}   & \textbf{A1}    & \textbf{A2}  & \textbf{A3}  & \textbf{(Enhanced)}  \\\hline
UART   & K64F & 947   & 31(119)      & 18                            & 44          & 28          & 64          & 21         & 5           & 90    \\
      & SAM3X & 588   & 18(175)      & 17                            & 64          & 25          & 82          & 24         & 0           & 106   \\
      & F103 & 331   & 7(56)        & 6                             & 8           & 13          & 14          & 11         & 2           & 27    \\
      & F429 & 317   & 7(52)        & 6                             & 8           & 12          & 14          & 10         & 2           & 26    \\
      & L152 & 331   & 7(56)        & 6                             & 8           & 13          & 14          & 11         & 2           & 27    \\
      \hdashline[1pt/5pt]
I2C    & K64F & 315   & 12(46)       & 5                             & 15          & 10          & 21          & 8          & 1           & 30    \\
      & SAM3X & 264   & 11(75)       & 12                            & 39          & 16          & 53          & 14         & 0           & 67    \\
      & F103 & 334   & 9(66)        & 14                            & 10          & 19          & 28          & 13         & 2           & 43 (5)    \\
      & F429 & 344   & 10(69)       & 14                            & 11          & 19          & 29          & 13         & 2           & 44 (5)    \\
      & L152 & 334   & 9(66)        & 14                            & 10          & 19          & 28          & 13         & 2           & 43 (5)    \\
      \hdashline[1pt/5pt]
SPI    & K64F & 551   & 11(90)       & 7                             & 24          & 11          & 34          & 6          & 2           & 42  (3)  \\
      & SAM3X & 254   & 11(58)       & 4                             & 28          & 7           & 32          & 7          & 0           & 39    \\
      & F103 & 264   & 9(49)        & 3                             & 10          & 5           & 14          & 2          & 2           & 18    \\
      & F429 & 269   & 9(50)        & 3                             & 10          & 5           & 14          & 2          & 2           & 18    \\
      & L152 & 264   & 9(49)        & 3                             & 10          & 5           & 14          & 2          & 2           & 18    \\
      \hdashline[1pt/5pt]
GPIO   & K64F & 50    & 6(6)         & 0                             & 9           & 0           & 9           & 0          & 0           & 9     \\
      & SAM3X & 1,502  & 43(1,245)     & 33                            & 539         & 31          & 572         & 31         & 0           & 603   \\
      & F103 & 176   & 14(39)       & 0                             & 14          & 0           & 14          & 0          & 0           & 14    \\
      & F429 & 73    & 10(13)       & 0                             & 12          & 0           & 12          & 0          & 0           & 12    \\
      & L152 & 79    & 11(14)       & 0                             & 13          & 0           & 13          & 0          & 0           & 13    \\
      \hdashline[1pt/5pt]     
TIMER    & K64F & 71    & 5(12)        & 4                             & 18          & 8           & 26          & 4          & 0           & 30    \\
      & SAM3X & 1045  & 41(197)      & 32                            & 95          & 27          & 127         & 27         & 0           & 154   \\
      & F103 & 576   & 20(115)      & 11                            & 30          & 9           & 41          & 9          & 0           & 50    \\
      & F429 & 410   & 19(93)       & 8                             & 14          & 7           & 22          & 7          & 0           & 29    \\
      & L152 & 576   & 20(115)      & 10                            & 41          & 7           & 51          & 7          & 0           & 58    \\
        \hdashline[1pt/5pt]
ETH    & F429 & 1,304  & 61(297)      & 7                             & 62          & 70          & 110         & 7          & 22          & 139 (3) \\
    \hdashline[1pt/5pt]

ADC    & K64F & 349   & 24(70)       & 9                             & 30          & 4           & 39          & 2          & 2           & 41    \\
       & SAM3X & 397   & 21(184)      & 70                            & 112         & 22          & 183         & 21         & 0           & 204  (3) \\
       & F103 & 325   & 14(78)       & 8                             & 31          & 4           & 39          & 2          & 2           & 43    \\
       & F429 & 506   & 17(112)      & 9                             & 37          & 5           & 46          & 3          & 2           & 51    \\
       & L152 & 583   & 20(137)      & 16                            & 39          & 6           & 56          & 3          & 2           & 61    \\
      \hdashline[1pt/5pt]
FTM  & K64F & 626   & 25(180)      & 12                            & 39          & 9           & 51          & 9          & 0           & 60    \\\hdashline[1pt/5pt]
PWM       & SAM3X & 324   & 43(242)      & 10                            & 192         & 72          & 202         & 72         & 0           & 274   \\
      \hdashline[1pt/5pt]
DAC    & SAM3X & 120   & 13(40)       & 5                             & 26          & 4           & 31          & 4          & 0           & 35    \\
       & L152 & 115   & 13(33)       & 0                             & 25          & 0           & 25          & 0          & 0           & 25    \\
      \hdashline[1pt/5pt]
MCG    & K64F & 417   & 12(48)       & 0                             & 13          & 8           & 21          & 0          & 0           & 21 (4)   \\
      \hdashline[1pt/5pt]
RCC    & F103 & 703   & 12(144)      & 17                            & 12          & 9           & 33          & 5          & 0           & 38  (6)  \\
       & F429 & 1,443  & 25(292)      & 21                            & 25          & 21          & 60          & 7          & 0           & 67  (6)   \\
       & L152 & 856   & 14(174)      & 19                            & 14          & 16          & 42          & 7          & 0           & 49   (6)  \\
      \hdashline[1pt/5pt]
PORT   & K64F & 1,868  & 38(359)      & 32                            & 38          & 96          & 134         & 32         & 0           & 166   \\
      \hdashline[1pt/5pt]
EXTI   & F103 & 104   & 6(56)        & 0                            & 6           & 40          & 6          & 40         & 0           & 46    \\
       & F429 & 104   & 6(56)        & 0                            & 6           & 46          & 6          & 46         & 0           & 52    \\
       & L152 &104   & 6(56)        & 0                            & 6           & 48          & 6          & 48         & 0           & 54    \\
      \hdashline[1pt/5pt]
PMC    & SAM3X & 288   & 26(223)      & 8                             & 142         & 9           & 159         & 0          & 0           & 159 (1)  \\
      \hdashline[1pt/5pt]
DMA    & F103 & 239   & 6(75)        & 0                             & 58          & 0           & 58          & 0          & 0           & 58    \\
       & F429 & 409   & 10(131)      & 8                             & 84          & 0           & 92          & 0          & 0           & 92    \\
       & L152 & 239   & 6(75)        & 0                             & 58          & 0           & 58          & 0          & 0           & 58    \\
      \hdashline[1pt/5pt]
DMA2D  & F429 & 175   & 22(70)       & 6                             & 22          & 5           & 33          & 0          & 0           & 33    \\
      \hdashline[1pt/5pt]
RTC    & K64F & 349   & 10(47)       & 0                             & 10          & 3           & 10          & 3          & 0           & 13    \\
      \hdashline[1pt/5pt]
SIM    & K64F & 573   & 22(118)      & 1                             & 22          & 0           & 23          & 0          & 0           & 23    \\
      \hdashline[1pt/5pt]
SMC    & K64F & 33    & 4(10)        & 0                             & 4           & 0           & 4           & 0          & 0           & 4     \\
      \hdashline[1pt/5pt]
WDOG   & K64F & 134   & 12(23)       & 0                             & 13          & 0           & 13          & 0          & 0           & 13    \\
       & SAM3X & 15    & 3(12)        & 2                             & 3           & 0           & 5           & 0          & 0           & 5     \\
       & F103 & 16    & 4(5)         & 2                             & 4           & 0           & 6           & 0          & 0           & 6     \\
      \hdashline[1pt/5pt]
CHIPID & SAM3X & 11    & 2(9)         & 0                             & 2           & 0           & 2           & 0          & 0           & 2     \\
      \hdashline[1pt/5pt]
EFC    & SAM3X & 21    & 4(10)        & 2                             & 8           & 2           & 10          & 2          & 0           & 12    \\
FLASH  & F103 & 121   & 9(36)        & 4                             & 10          & 2           & 14          & 2          & 0           & 16    \\
       & F429 & 123   & 9(36)        & 5                             & 9           & 2           & 14          & 2          & 0           & 16    \\
       & L152 & 40    & 12(15)       & 3                             & 12          & 0           & 15          & 0          & 0           & 15    \\
      \hdashline[1pt/5pt]
PWR    & F429 & 42    & 2(22)        & 4                             & 2           & 0           & 6           & 0          & 0           & 6     \\
       & L152 & 51    & 2(25)        & 7                             & 2           & 0           & 9           & 0          & 0           & 9     \\
      \hdashline[1pt/5pt]
SYSCFG & F429 & 32    & 7(11)        & 1                             & 7           & 0           & 8           & 0          & 0           & 8     \\    
\hline  
\end{tabular}
\end{adjustbox}
\vspace*{-6mm}
\end{table}

\section{Detail of Additional Samples}
\label{app:addtests}

We detail the additional samples used in enhanced \ca rule evaluation in Section~\ref{sec:enhance}.

\paragraph{PinLock} is a piece of firmware for smart lock which has also been used in evaluation in related work~\cite{clements2018aces}. It reads a PIN number through a UART interface, hashes it to compare with a known hash, and sends a signal to an I/O pin to unlock a digital lock when the PIN
is correct. 

\paragraph{Shell} is shipped with NXP K64F SDK~\cite{k64fsdk}.
It shows a text-based UI interface that accepts commands
to control four LEDs.


\paragraph{SPI\_FullDuplex} is included in the STMicroelectronics F4xx chip SDK~\cite{stm32f4sdk}. The original demo includes two separate firmware
that runs on two boards for sending and receiving via SPI. We modified it to run on a single broad. Specifically, we use SPI1 as master to send data, and SPI2 as slave to receive data.
The firmware also checks whether the received data is the same as the sent one.
Both SPI master and slave modes are tested.

\paragraph{AnalogWatchdog} monitors the amplitude of the input signal of the ADC analog channel. If a threshold is exceeded, an error handler will be invoked.
It uses GPIO, ADC, Timer and DMA on STM32F4xx chips~\cite{RMSTM32L132}.
Specifically, GPIO is used in analog mode to drive signal from device pin to ADC input and the timer is used to trigger ADC conversions.
The conversion results are transferred automatically by DMA.

\paragraph{Others} include nine demos from chip vendor SDKs used for ADC, SPI and UART data transmission in DMA mode on F103, F429 and L152, and four demos for TCP and UDP communication over Ethernet on F429.



\section{QEMU Peripheral Model Implementations}
Listing~\ref{lst:QEMUUART} and ~\ref{lst:QEMUADC} show
code snippets of the QEMU peripheral model implementations
for STMF405/205 UART and ADC, respectively.

\begin{lstlisting}[style=CStyle,label={lst:QEMUUART},xleftmargin=1em,framexleftmargin=1em,frame=shadowbox,caption={QEMU backend for STMF405/205 UART and annotated \ca rules}]
static void stm32f2xx_usart_receive(void *opaque, const uint8_t *buf, int size)
{
    ...
    s->usart_dr = *buf;      
    s->usart_sr |= USART_SR_RXNE; //C-A Rule: B #DR[R] > 0 -> SR[RXNE] := 1
    if (s->usart_cr1 & USART_CR1_RXNEIE) {
        qemu_set_irq(s->irq, 1);//C-A Rule: V SR[RXNE] == 1 & V CR1[RXNEIE] == 1 -> IRQ[RXNE]:= Ready 
    }

}
static void stm32f2xx_usart_reset(DeviceState *dev) {
        STM32F2XXUsartState *s = STM32F2XX_USART(dev);
        /* Setting reset values */
        s->usart_sr = USART_SR_RESET;
        ....
}
static uint64_t stm32f2xx_usart_read(void *opaque, hwaddr addr, unsigned int size)
{
    ...
    switch (addr) {
    case USART_DR:
        retvalue = s->usart_dr & 0x3FF;
        s->usart_sr &= ~USART_SR_RXNE;//C-A Rule: B #DR[T] <= 0 -> SR[RXNE] := 0
        qemu_chr_fe_accept_input(&s->chr);
        qemu_set_irq(s->irq, 0);//C-A Rule: V SR[RXNE] == 0 & V CR1[RXNEIE] == 1 -> IRQ[RXNE]:= Enable       
        return retvalue;
        ...
    }

    return 0;
}
static void stm32f2xx_usart_write(void *opaque, hwaddr addr, uint64_t val64, unsigned int size)
{
    ...
    switch (addr) {
    case USART_SR:
        if (value <= 0x3FF) {
            s->usart_sr = value | USART_SR_TXE;//C-A Rule: O* -> SR[TXE]:= 1
        } else {
            s->usart_sr &= value;
        }
        if (!(s->usart_sr & USART_SR_RXNE)) {
            qemu_set_irq(s->irq, 0);//C-A Rule: V SR[RXNE] == 0 & V CR1[RXNEIE] == 1->IRQ[RXNE]:= Enable 
        }
        return;
    case USART_DR:
        if (value < 0xF000) {
            ch = value;
            qemu_chr_fe_write_all(&s->chr, &ch, 1);
            s->usart_sr |= USART_SR_TC;// C-A Rule: #D[T] <= 0 -> SR[TC]:= 1
        }
        return;
        ...
    }
}
....
\end{lstlisting}

\begin{lstlisting}[style=CStyle,label={lst:QEMUADC},xleftmargin=1em,framexleftmargin=1em,frame=shadowbox,caption={QEMU backend for STMF405/205 ADC which shows logic that \sys failed to extract from the manual}]
static uint32_t stm32f2xx_adc_generate_value(STM32F2XXADCState *s)
{
    s->adc_dr = s->adc_dr + 7;

    switch ((s->adc_cr1 & ADC_CR1_RES) >> 24) {
    case 0:
        /* 12-bit */
        s->adc_dr &= 0xFFF;
        break;
    case 1:
        /* 10-bit */
        s->adc_dr &= 0x3FF;
        break;
    case 2:
        /* 8-bit */
        s->adc_dr &= 0xFF;
        break;
    default:
        /* 6-bit */
        s->adc_dr &= 0x3F;
    }

    if (s->adc_cr2 & ADC_CR2_ALIGN) {
        /* Left alignment */
        return (s->adc_dr << 1) & 0xFFF0;
    } else {
        /* Right alignment */
        return s->adc_dr;
    }
}
\end{lstlisting}

\section{Trace Fidelity Score}
Table~\ref{tab:score} shows the results of
trace fidelity scores.

\begin{table}[!ht]
\setlength\abovedisplayskip{0pt}
\setlength\belowdisplayskip{0pt}
\caption{Trace Fidelity Score}
\vspace{-3mm}
\centering
\begin{adjustbox}{max width=0.9\columnwidth}

\setlength\abovedisplayskip{0pt}
\setlength\belowdisplayskip{0pt}
\begin{tabular}{lrrrrrrrr}

\hline
\multicolumn{9}{c}{\textbf{Initialization}}                                                                                                                                                                                                                                                                                                               \\ \hline
\multicolumn{1}{c|}{}                             & \multicolumn{2}{c}{\textbf{I2C}}                                        & \multicolumn{2}{c}{\textbf{UART}}                                       & \multicolumn{2}{c}{\textbf{GPIO}}                                       & \multicolumn{2}{c}{\textbf{TIMER}}                                      \\ \hline
\multicolumn{1}{c|}{}                             & K64F                               & F103                               & K64F                               & F103                               & K64F                               & F103                               & K64F                               & F103                               \\
\multicolumn{1}{c|}{\ppim}         & 100\%                              & 26\%                               & 100\%                              & 100\%                              & 100\%                              & 100\%                              & 100\%                              & 100\%                              \\
\multicolumn{1}{c|}{\uemu}         & 100\%                              & 100\%                              & 100\%                              & 100\%                              & 99\%                               & 100\%                              & 100\%                              & 100\%                              \\
\multicolumn{1}{c|}{\textbf{\sys}} & \textbf{100\%}                     & \textbf{100\%}                     & \textbf{100\%}                     & \textbf{100\%}                     & \textbf{100\%}                     & \textbf{100\%}                     & \textbf{100\%}                     & \textbf{100\%}                     \\ \hline
\multicolumn{9}{c}{\textbf{Main Loop}}                                                                                                                                                                                                                                                                                                                    \\ \hline
\multicolumn{1}{c|}{}                             & \multicolumn{2}{c}{\textbf{I2C}}                                        & \multicolumn{2}{c}{\textbf{UART}}                                       & \multicolumn{2}{c}{\textbf{GPIO}}                                       & \multicolumn{2}{c}{\textbf{TIMER}}                                      \\ \hline
\multicolumn{1}{c|}{}                             & K64F                               & F103                               & K64F                               & F103                               & K64F                               & F103                               & K64F                               & F103                               \\
\multicolumn{1}{c|}{\ppim}         & 10\%                               & 9\%                                & 43\%                               & 59\%                               & 100\%                              & 100\%                              & 100\%                              & 100\%                              \\
\multicolumn{1}{c|}{\uemu}         & 91\%                               & 21\%                               & 88\%                               & 98\%                               & 100\%                              & 100\%                              & 100\%                              & 100\%                              \\
\multicolumn{1}{c|}{\textbf{\sys}} & \textbf{100\%}                     & \textbf{100\%}                     & \textbf{100\%}                     & \textbf{100\%}                     & \textbf{100\%}                     & \textbf{100\%}                     & \textbf{100\%}                     & \textbf{100\%}                     \\ \hline
\multicolumn{9}{c}{\textbf{Interrupt}}                                                                                                                                                                                                                                                                                                                    \\ \hline
\multicolumn{1}{c|}{}                             & \multicolumn{2}{c}{\textbf{I2C}}                                        & \multicolumn{2}{c}{\textbf{UART}}                                       & \multicolumn{2}{c}{\textbf{GPIO}}                                       & \multicolumn{2}{c}{\textbf{TIMER}}                                      \\ \hline
\multicolumn{1}{c|}{}                             & K64F                               & F103*                               & K64F                               & F103                               & K64F                               & F103                               & K64F                               & F103                               \\
\multicolumn{1}{c|}{\ppim}         & 11\%                               & -                              & 10\%                               & 10\%                               & 4\%                                & 9\%                                & 29\%                               & 85\%                               \\
\multicolumn{1}{c|}{\uemu}         & 5\%                                & -                              & 40\%                               & 39\%                               & 15\%                               & 47\%                               & 86\%                               & 92\%                               \\
\multicolumn{1}{c|}{\textbf{\sys}} & \textbf{100\%}                     & \textbf{-}                     & \textbf{100\%}                     & \textbf{100\%}                     & \textbf{100\%}                     & \textbf{100\%}                     & \textbf{100\%}                     & \textbf{100\%}                     \\ \hline
\multicolumn{9}{c}{\textbf{Combined}}                                                                                                                                                                                                                                                                                                                      \\ \hline
\multicolumn{1}{c|}{}                             & \multicolumn{2}{c}{\textbf{I2C}}                                        & \multicolumn{2}{c}{\textbf{UART}}                                       & \multicolumn{2}{c}{\textbf{GPIO}}                                       & \multicolumn{2}{c}{\textbf{TIMER}}                                      \\ \hline
\multicolumn{1}{c|}{}                             & K64F                               & F103                               & K64F                               & F103                               & K64F                               & F103                               & K64F                               & F103                               \\
\multicolumn{1}{c|}{\ppim}         & \multicolumn{1}{r}{40\%}           & \multicolumn{1}{r}{17\%}           & \multicolumn{1}{r}{51\%}           & \multicolumn{1}{r}{56\%}           & \multicolumn{1}{r}{68\%}           & \multicolumn{1}{r}{70\%}           & \multicolumn{1}{r}{76\%}           & \multicolumn{1}{r}{95\%}           \\
\multicolumn{1}{c|}{\uemu}         & \multicolumn{1}{r}{66\%}           & \multicolumn{1}{r}{61\%}           & \multicolumn{1}{r}{76\%}           & \multicolumn{1}{r}{79\%}           & \multicolumn{1}{r}{71\%}           & \multicolumn{1}{r}{82\%}           & \multicolumn{1}{r}{95\%}           & \multicolumn{1}{r}{97\%}           \\
\multicolumn{1}{c|}{\textbf{\sys}} & \multicolumn{1}{r}{\textbf{100\%}} & \multicolumn{1}{r}{\textbf{100\%}} & \multicolumn{1}{r}{\textbf{100\%}} & \multicolumn{1}{r}{\textbf{100\%}} & \multicolumn{1}{r}{\textbf{100\%}} & \multicolumn{1}{r}{\textbf{100\%}} & \multicolumn{1}{r}{\textbf{100\%}} & \multicolumn{1}{r}{\textbf{100\%}} \\ \hline
\end{tabular}
\label{tab:score}
\end{adjustbox}
\flushleft
\scriptsize{
*: The F103 I2C unit-test sample does not involve any external interrupts.
Note that
all three emulators have the 100\% score with the ADC and SPI unit test samples which are not listed in the Table.
}
\vspace{-6mm}
\end{table}

\clearpage

\end{document}